# Fusion cross sections for $^{6,7}Li + ^{24}Mg$ reactions at energies below and above the barrier

M. Ray[1], A. Mukherjee[2], M. K. Pradhan[2], Ritesh Kshetri[2], M. Saha Sarkar[2], R. Palit[3], I. Majumdar[3], P. K. Joshi[3], H. C. Jain[3] and B. Dasmahapatra[2*]

[1.] *Behala College, Parnasree, Kolkata – 700060, India*

[2.] *Saha Institute of Nuclear Physics, 1/AF, Bidhannagar, Kolkata – 700064, India*

[3.] *Tata Institute of Fundamental Research, Mumbai – 400005, India*

**Abstract**: Measurement of fusion cross sections for the $^{6,7}Li + ^{24}Mg$ reactions by the characteristic γ -ray method has been done at energies from below to well above the respective Coulomb barriers. The fusion cross sections obtained from these γ-ray cross sections for the two systems are found to agree well with the total reaction cross sections at low energies. The decrease of fusion cross sections with increase of energy is consistent with the fact that other channels, in particular breakup open up with increase of bombarding energy. This shows that there is neither inhibition nor enhancement of fusion cross sections for these systems at above or below the barrier. The critical angular momenta ($l_{cr}$) deduced from the fusion cross sections are found to have an energy dependence similar to other Li - induced reactions.

[*] Corresponding author's e-mail: binay.dasmahapatra@saha.ac.in

# I. INTRODUCTION

Investigation of fusion reactions induced by weakly bound projectiles at energies close to the Coulomb barrier is a field of great interest over the last few years. This has primarily been motivated by the present availability of light radioactive (loosely bound) ion beams, some of which exhibit unusual features like halo/skin structure and very small binding energy of the last nucleon(s). Measurement of fusion cross sections for the systems containing such nuclei is interesting in view of the fact that one may expect to observe either enhanced fusion cross sections because of the larger spatial extent of such nuclei or inhibition of the same due to their greater probability for breakup into two or more constituents because of their low binding energies.

Because of low intensity and poor energy resolution of radioactive ion beams, measurement of fusion cross section involving them is still difficult, though few measurements of fusion cross sections have been reported very recently [1-4] . On the other hand, it is very convenient to produce high intensity stable beams (of $^{9}$Be, $^{6}$Li and $^{7}$Li) that are weakly bound and consequently should have a significant breakup probability. Though there have been many theoretical and experimental works on this subject, the reaction mechanism is still far from being well understood [5]. A full understanding of fusion and breakup processes induced by these loosely bound nuclei may serve as an important reference for similar studies involving radioactive nuclei.

Fusion cross section measurement of the reactions involving heavy target masses and loosely bound stable projectiles $^{9}$Be + $^{144}$Sm [5]; $^{9}$Be + $^{208}$Pb [6]; $^{9}$Be + $^{209}$Bi [7]; $^{6,7}$Li + $^{209}$Bi [8]; $^{7}$Li + $^{165}$Ho [9] and $^{7}$Li, $^{10}$B + $^{159}$Tb [10] show suppression in the complete fusion cross sections at energies above their respective Coulomb barriers ($V_b$) when compared with the prediction of one dimensional BPM (Barrier Penetration Model). For medium and light mass nuclei, owing to the experimental difficulties, only total (complete + incomplete) fusion cross sections have been measured for the systems like $^{6,7}$Li, $^{9}$Be + $^{64}$Zn [11,12]; $^{6,7}$Li + $^{59}$Co [13]; $^{6,7}$Li, $^{7,9}$Be + $^{27}$Al [14-17]; $^{6,7}$Li + $^{16}$O[18,19]; $^{6,7}$Li + $^{12,13}$C [20-22]. These measurements do not show any suppression of total fusion cross sections at above-barrier energies. It may be mentioned that fusion cross sections for the systems mentioned above were measured using either characteristic γ-ray yield method or evaporation residue detection technique depending on the systems and energy regime of interest.

Considering the present scenario of target mass dependence of fusion cross sections we planned to measure fusion cross section for the $^{6,7}Li + ^{24}Mg$ systems covering a wide energy range from below to substantially (~ 3 times) above the respective Coulomb barriers using the characteristic γ-ray technique. Usually the γ-ray method is used for the measurement of fusion cross sections for the systems at low energies [19-21, 23-25]. However, the method can be extended to the higher energies for the systems where the γ-ray yield is not very low [25-28]. It may further be mentioned that so far there has been no fusion or total reaction cross section measurement for these systems.

# II. THE EXPERIMENTAL SETUP AND MEASUREMENT

The energy level diagrams for $^{6}Li + ^{24}Mg$ and $^{7}Li + ^{24}Mg$ reactions are shown in Fig.1 and 2 respectively. They illustrate the expected channels, the residues and the de-exciting γ-rays from the residues following the two reactions.

The measurements were performed using the 3MV Pelletron accelerator at Institute of Physics (IOP), Bhubaneswar and 14UD BARC-TIFR Pelletron Accelerator Facility at Tata Institute of Fundamental Research (TIFR), Mumbai. The energy ranges covered in the two accelerator centres are $E_{lab}$= 6.0 – 11.5 MeV at IOP and 11.0 – 30.0 MeV at TIFR, respectively. At IOP natural Mg – foil (316 ± 72.0 $\mu g/cm^2$) backed by a solid tantalum sheet was used. The target was placed in a specially designed scattering chamber. It consists of two concentric stainless steel cylinders insulated from each other. The 25mm diameter inner cylinder has the provision of holding the experimental target at one of its ends. This inner cylinder together with the target constitutes the Faraday cup measuring the total charge. Beams of $^{6}Li^{2+,3+}$ and $^{7}Li^{2+,3+}$ were used and the current varied between 10-40nA. The γ-rays produced during the bombardment were detected by an HPGe detector of ~ $60cm^3$ volume, placed at 55° with target-to-detector distance ≈ 6.5cm. Both on-line and off-line spectra were taken for each exposure. The off-line spectra enabled us to identify the normal background γ-ray lines or the activity lines arising due to beam bombardments. The measurements were done in steps of ~1MeV for both the reactions. The energy range covered corresponds to $E_{c.m.}$= 4.62 – 8.84 MeV for $^{6}$Li + $^{24}$Mg and $E_{c.m.}$= 5.23 – 8.22 MeV for $^{7}$Li + $^{24}$Mg reactions after making necessary correction for energy loss in the target.

At TIFR, a self supporting target of natural Mg (1.24 ± 0.05 mg/cm$^{2}$) was put at the centre of an 80mm diameter reaction chamber. The characteristic γ-rays emitted by the fusion evaporation residues were detected with a Compton suppressed clover detector placed at 55° with a target-to-detector distance ~12cm. The total charge of each exposure was measured in a 30cm long tube, insulated from the chamber, serving as the Faraday Cup. Beam current was varied between 2-10nA. The measurements were done in steps of ~1-2MeV. The energy range covered corresponds to $E_{c.m.}$= 8.3 – 23.7 MeV for $^{6}$Li + $^{24}$Mg and $E_{c.m.}$= 7.92 – 22.9 MeV for $^{7}$Li + $^{24}$Mg reactions respectively after making necessary correction for energy loss in the target.

Thicknesses of the targets were determined by weighing the rolled target sheets and measuring their areas. The magnesium targets used here were prepared from natural magnesium material (79% $^{24}$Mg, 10% $^{25}$Mg and 11% $^{26}$Mg). In course of analysis of the spectra, possible interference of γ-rays from $^{25}$Mg and $^{26}$Mg in the $^{6,7}$Li + $^{24}$Mg reaction has also been considered. From the yield of the characteristic γ-rays following the reactions with $^{25,26}$Mg, it has been found that their contributions as contaminants in fusion cross sections of the $^{6,7}$Li + $^{24}$Mg reactions are negligible. In addition to the spectra obtained with the target, spectra with beam on a Ta-sheet (used as backing) and a Ta-frame having a hole in place of the target were also obtained. These spectra together with beam-off background spectra enabled us to identify the impurity lines and subtract the contribution if necessary.

# III. ANALYSIS AND RESULTS

## 1. The γ-ray cross sections

Typical γ-ray spectra of the two reactions at $E_{lab}$=11MeV obtained at IOP, Bhubaneswar are shown in Fig.3. In this figure the γ-rays originating from the residual nuclei are marked by square brackets, while the "background" γ-rays as mentioned above are marked by alphabets and identified in Table1.

The γ-ray cross sections ($\sigma_\gamma$) were obtained from the relation

$$\sigma_\gamma = \frac{N_\gamma}{\varepsilon_\gamma N_B N_T} \qquad (1)$$

where, $N_\gamma$ is the number of counts under the γ-ray peak, $\varepsilon_\gamma$ is the absolute full energy peak detection efficiency of the detector for the specific γ-ray. $N_B$ and $N_T$ are the number of beam particles and number of target nuclei, respectively. The procedure for the measurement of $N_B$, $N_T$ and $\varepsilon_\gamma$ has been described in details in an earlier work [20]. The total systematic uncertainty in the γ-ray cross section measurement is found to be ~ 11%.

## 2. The channel cross sections

A characteristic γ-ray is emitted from a residual nucleus in the reaction process when the excited state from which the emission occurs is populated either directly from particle evaporations or via γ-ray cascades originating in the higher states, below the particle emission threshold of the residual nucleus. To extract the channel cross sections one needs the "branching factor", $f_\gamma = \sigma_\gamma / \sigma_{ch}$, giving the fraction of the residual nuclei emitting the characteristic γ-ray when left in the bound states. For finding $f_\gamma$, one needs, for the nucleus under consideration, the relative population of different bound states as well as their branching ratios. While the branching ratios can be obtained from the known de-excitation schemes of the nuclei, the relative populations of bound states must be evaluated by a statistical model calculation. The procedure of finding $f_\gamma$ has been described earlier [20]. Of the characteristic γ-rays only a few (which were found to be contaminant free and intense) were used to determine the channel cross sections. The relevant ‘branching factors’ calculated with the statistical model code CASCADE [29] have been shown in Fig.4(a) and Fig.4(b) corresponding to the reactions $^{6}$Li + $^{24}$Mg and $^{7}$Li + $^{24}$Mg respectively. The cross sections for some of the channels thus determined from the measured γ-ray cross sections using these $f_\gamma$ for the two reactions are shown in Fig.5 and Fig.6 respectively. For comparison, the channel cross sections calculated using the code CASCADE are also shown in the same figures.

We now discuss certain relevant points in the determination of some of the channel cross sections.

### A. pn exit channel

$^{28}$Si + pn channel of $^{6}$Li + $^{24}$Mg reaction and $^{29}$Si + pn channel of $^{7}$Li + $^{24}$Mg reaction constitute about 50-70% and 35-70% respectively (as per CASCADE calculation) of the total reaction cross sections in the energy range ($E_{lab}$~6–30MeV) of our investigation. Hence measurement of the cross sections for the pn channel is very important in the determination of fusion cross sections for the two systems. The γ-ray peaks at 1.779MeV and 1.273MeV corresponding to the first excited state to ground state transitions of the residual nuclei $^{28}$Si and $^{29}$Si respectively are found to be quite distinct and pose no difficulties in finding out their areas at low energies. But at higher bombarding energies the 1.779MeV γ-ray was found to be contaminated with that originating from the $\beta^-$ decay of $^{28}$Al (2.24m) [$^{28}$Al+2p channel] and the shape of the peak got distorted more and more. Hence the $^{28}$Si + pn channel cross sections at higher incident energies were obtained from the contaminant free 2.837MeV (4.617MeV → 1.779MeV) γ-ray of $^{28}$Si though it’s intensity was rather small. At low bombarding energies (at IOP) where the channel cross section could be obtained using either the 1.779MeV γ-ray peak or the 2.837 MeV γ-ray peak, it was found that the measured values of channel cross sections are consistent with each other.

The cross sections for the $^{29}$Si + pn channel of $^{7}$Li + $^{24}$Mg reaction on the other hand could be determined from the 1.273 MeV γ-ray for the entire region except for a few at high bombarding energies, where because of population of the $^{29}$Al+2p channel the area of the peak was difficult to determine by separating it from the contaminant peak (1.273MeV) arising due to the $\beta^-$ decay of $^{29}$Al (6.6m) to $^{29}$Si. Nevertheless the cross sections for the same channel ($^{29}$Si + pn) could be determined from two more characteristic γ-rays of $^{29}$Si, namely 1.596MeV (3.624MeV → 2.028MeV) and 2.028MeV (2.028MeV → 0.0MeV) and these γ-rays together with the 1.273MeV γ-ray yield consistent channel cross sections (Fig.6).

**B. pα exit channel**

The $^{25}$Mg + pα channel of $^{6}$Li + $^{24}$Mg reaction is quite prominent even at low bombarding energies but this is not the case with the $^{26}$Mg + pα channel of $^{7}$Li + $^{24}$Mg reaction. Out of the prominent characteristic γ-rays of $^{25}$Mg: 0.585MeV (0.585MeV → 0), 0.975MeV (0.975MeV → 0) and 0.390MeV (0.975MeV → 0.585MeV), the 0.585 MeV γ-ray yields much larger cross sections compared to those obtained from the other two γ-rays (Fig.5). The cross sections for the 0.585MeV γ-ray remained practically unaltered even after the subtraction of the contribution from the 0.583MeV background radioactive line and the 0.583MeV γ-ray of $^{22}$Na [2α + $^{22}$Na channel] estimated from the background spectra and 0.891MeV γ-ray cross section of $^{22}$Na (obtained at relatively higher bombarding energies) respectively. The situation appears to be similar to $^{6}$Li + $^{12}$C reaction [21,23] where also the cross sections for the 3.089MeV γ-ray of $^{13}$C showed a very large cross section compared to other γ-rays of the same nucleus. In view of these, the cross sections for the $^{25}$Mg + αp channel were obtained from the 0.975MeV and 0.390MeV γ-rays and are found to agree well with the Statistical Compound Nucleus (SCN) calculations over a wide range of energy. For the $^{26}$Mg + pα channel of $^{7}$Li + $^{24}$Mg reaction, on the other hand, cross sections determined from the 1.808 MeV γ-ray peak corresponding to the first excited state to the ground state transition of $^{26}$Mg, show good agreement with the SCN calculations at low bombarding energies. At high bombarding energies, however, the peak corresponding to this γ-ray could not be separated from the dominant 1.779MeV γ-ray peak corresponding to $^{28}$Si.

**C. pnα exit channel**

The three particle evaporation channels corresponding to the $^{24}$Mg + pnα channel of $^{6}$Li + $^{24}$Mg reaction and the $^{25}$Mg + pnα channel of $^{7}$Li + $^{24}$Mg reaction are found to contribute significantly at higher bombarding energies. In contrast to the excitation of

$^{25}$Mg in the $^{7}$Li + $^{24}$Mg reaction the characteristic γ-ray 1.368MeV of $^{24}$Mg in the $^{6}$Li + $^{24}$Mg reaction is observable at very low bombarding energies (Fig.3). As the emissions of three particles are expected only at very high bombarding energies which is also corroborated by the CASCADE calculations, such excitation of $^{24}$Mg at low bombarding energies must be due to some other processes like $^{24}$Mg (n,n′γ) and are not considered in the evaluation of channel or fusion cross sections.

**D. 2pn exit channel**

Besides αpn – three particle emission channel, the 2pn + $^{27}$Al channel of $^{6}$Li+$^{24}$Mg reaction appears to contribute significantly at very high bombarding energies. However, the characteristic γ-rays of $^{27}$Al, especially the 0.844MeV γ-ray peak is observed even at the lowest bombarding energies for both the reactions. The excitation of $^{27}$Al like $^{24}$Mg is also attributed to be due to the $^{27}$Al (n,n′γ) reaction. The 0.844MeV γ-ray peak is further contaminated by the 0.847MeV γ-ray of $^{56}$Fe (n,n′γ) reaction. At bombarding energies above 25MeV ($E_{c.m.}$~ 20MeV) the γ-ray spectra for the $^{6}$Li+$^{24}$Mg reaction are observed to be dominated by the γ-rays of $^{24}$Mg and $^{27}$Al only. Only at these energies where the contribution of the (n,n′γ) reaction is significantly smaller than that due to the reaction $^{6}$Li + $^{24}$Mg → 2pn + $^{27}$Al, the contribution of $^{27}$Al+2pn channel was determined from the 1.014MeV γ-ray of $^{27}$Al.

## *3.A. <u>Total fusion cross sections from the sum of the cross sections for the exit channels.</u>*

The conventional way to determine the total fusion cross sections for any reaction is to sum the exit channel cross sections which are believed to be due to de-excitation of the compound nucleus. As the cross sections for most of the exit channels of the two reactions could be determined, we sum them to get the total fusion cross sections for the two reactions. It should, however, be mentioned that for the channels pn+$^{28}$Si and 2p+$^{28}$Al $\xrightarrow{\beta^-}$ $^{28}$Si of the $^{6}$Li+$^{24}$Mg reaction where the characteristic γ-ray (1.779MeV) is same for both the channels, we determined the total channel cross sections (Fig.5) using the composite area and the relevant $f_\gamma$. For the $^{7}$Li+$^{24}$Mg reaction the peak corresponding to 1.779MeV γ-ray is further contaminated by the 1.808MeV γ-ray of $^{26}Mg$ and could not be separated at relatively higher bombarding energies. As a result we could determine the cross sections for the three channels together. This is shown in Fig.6.

The total fusion cross sections thus determined are shown in Fig.7 and Fig.8.

## *B. The total fusion cross sections from the sum of the cross sections for the γ-rays*

According to statistical model calculations as the de-excitation γ-rays of the residual nuclei originate from the compound nucleus formation, the total fusion cross section, in principle, could be determined from the cross section for any individual γ-ray. This was, in fact, shown in case the of $^{12}C+^{13}C$ reaction where the total fusion cross sections for the reaction were obtained separately from the cross sections for the three γ-rays of the residual nuclei [25]. However, when the cross section for an individual γ-ray is small the general practice is to take into account as many γ-rays as possible (which do not show any abnormal behaviour in their excitation functions), sum their cross sections and use a total $F_\gamma$ which corresponds to the ratio of total γ-ray cross sections and total channel cross sections (total fusion cross sections) both evaluated by the statistical model calculations (CASCADE) using the branching ratios from the literature. This method was used earlier by Scholz et. al. [24] in the determination of total fusion cross sections for the $^{7}Li+^{16}O$ reaction. It was, however, extensively used by us [18,21,30] in the determination of total fusion cross sections for $^{6}Li+^{12}C$, $^{6}Li+^{16}O$ and $^{7}Li+^{16}O$ reactions. The total fusion cross sections determined by this procedure are also shown in Fig.7 and Fig.8.

The cross sections obtained from the two procedures agree well with each other and they compare well with the total reaction cross sections obtained from the optical model calculations using parameters of the optical model potentials for the $^{6,7}Li+^{28}Si$ reactions [31] after proper scaling of mass. The dependence of $f_\gamma$ (or $F_\gamma$) on various parameters of the calculation using the code CASCADE has been studied by several authors including us. It is found from the detailed study that except for very weak γ-rays this correction factor is rather insensitive to the reasonable variation of these parameters and the uncertainty in it is estimated to be ≤ 10% [32]. The uncertainty in the correction factor (10%) has been added in quadrature to the total systematic uncertainty (~ 11%) in the experimental γ-ray cross sections resulting in ~ 15% uncertainty in the total fusion cross section.

### ***4. Angular momentum and fusion cross section***

The maximum angular momentum associated with the fusion process, commonly called the critical angular momentum, $l_{cr,}$ can be extracted from the measured fusion data using the well-known expression

$$\sigma_{fus} = \frac{\pi}{\kappa^2}\sum_{l=0}^{lcr}(2l+1) = \frac{\pi}{\kappa^2}(l_{cr}+1)^2 \qquad (2)$$

according to the sharp-cutoff approximation. These critical angular momenta, extracted from the fusion data, are shown in Fig.9. as a function of the compound nucleus excitation energy for both the above systems. This figure also shows the energy dependence of the grazing angular momenta, $l_{gr}$, for such systems (solid lines), calculated using the parameters of the optical model. It is found that the $l_{cr}$ values remain close to the $l_{gr}$ values at low bombarding energies and start diverging away from the $l_{gr}$ values thus indicating a limitation of fusion cross section. Such a behaviour of angular momentum has been observed for systems like $^{9}$Be+$^{9}$Be, $^{6,7}$Li+$^{12,13}$C and $^{6,7}$Li+$^{16}$O reactions [18,20,21,23].

## IV. DISCUSSIONS

Fig.10. shows the measured total fusion cross sections obtained as the average of “sum of channel cross sections” and fusion cross sections from the “sum of γ-ray cross sections” as described in Sec.III.

These cross sections are compared with the Optical Model (OM) calculations. Such calculations with parameters of the potential obtained from fitting of the elastic scattering data for a system are expected to yield the total reaction cross sections for the same system. In view of the lack of $^{6,7}$Li+$^{24}$Mg elastic scattering data we used the parameters of the potential from the data of nearby system $^{6,7}$Li+$^{28}$Si [31], after proper scaling due to the change of mass of the target.

It is observed that the measured cross sections are nearly equal to the total reaction cross sections at lower energies and start decreasing with the increase of bombarding energies. This observation is consistent with our previous measurement of fusion cross section for $^{6,7}$Li with light mass targets [18,20,21]. The decrease of measured cross sections compared to the total reaction cross section values at high bombarding energies appears to be natural since the quasi elastic channels other than fusion gradually open up with increase of incident energy. The measurement of cross sections at very high energy [$E_{lab}$ ~ 36MeV] shows that the total reaction cross section for a number of $^{6,7}$Li induced reactions is almost equal to the sum of fusion and break up cross sections [33,34]. As we do not find any exclusive evidence for neutron or $\alpha$-transfer reaction, it appears that the breakup process is the dominant quasi-elastic reaction at high energy region of the present measurement.

Considering the success of coupled channels calculations in medium and heavy systems one may attempt to do the same for the present two systems. The most appropriate potential for such calculations would have been the one obtained from fitting both elastic and fusion data for the systems. Such a potential not being available, we have used the potential recently used by Sinha et al. [35] for the $^{7}$Li + $^{28}$Si system derived from Anjos et al. [36] for the system $^{11}$B + $^{27}$Al. The calculations with CCFULL code [37] in no coupling mode (one dimensional barrier penetration model) for the present two systems are shown in the same figure (Fig10). The results do not agree with the fusion cross sections at sub-barrier energies. The theoretical cross sections remain practically unaltered on inclusion of coupling to excited states of $^{24}$Mg. Perhaps one could attempt a CDCC (continuum discretized coupled channels) calculations [38] to explain this enhancement. For such calculations however, it is necessary to have an appropriate potential and hence elastic scattering data for $^{6,7}$Li+$^{24}$Mg systems are required at the energies where fusion cross sections are measured.

In order to understand the behaviour of the $^{6,7}$Li+$^{24}$Mg reactions, we now compare their fusion cross sections with those of nearby $^{6,7}$Li+$^{27}$Al and $^{6,7}$Li+$^{28}$Si systems measured by different authors [15,17,35] (Fig11). It is observed that within experimental uncertainty the measured values for all the systems appear to be almost equal in this high energy region. This shows that the fusion cross sections for these systems are determined mainly by the gross properties of the colliding nuclei (e.g. charge, mass, radius etc.) which vary little from one system to another and the cross sections are practically independent of the microscopic properties (e.g. cluster character, valence nucleons etc.) of the interacting nuclei. It is to be noted that except for the $^{6,7}$Li+$^{24}$Mg reactions of the present work, there are no measurements for the other systems at lower energies (below barrier). Investigations of fusion cross sections for the light systems, in general, show that although the nature of energy dependence of the cross sections at higher energy region is similar, some systems behave very much differently at energies near and below the barrier [39,40]. Thus measurement of fusion cross sections for the above systems at low bombarding energies may prove to be interesting.

The lack of low energy fusion data may be complemented by the $^{6,7}$Li-induced reactions with the light mass targets $^{12,13}$C and $^{16}$O investigated earlier [18,20,21]. However, as the mass of these target nuclei ($^{12}$C, $^{13}$C, $^{16}$O) are much less than $^{24}$Mg, $^{27}$Al and $^{28}$Si, it seems more appropriate to plot the reduced cross sections ($\sigma_{fus}$ / $R_B^2$) as a function of $E_{c.m.}/V_B$ to account for the change of mass and barrier energy. The barrier parameters $R_B$ and $V_B$ are taken from the systematics proposed by Vaz et.al. [41]. The reduced fusion cross sections for the $^6$Li and $^7$Li induced reactions are shown in Fig.12 and Fig.13 respectively. The $^6$Li-induced reactions data appear to be more scattered than the $^7$Li induced reactions data especially near the barrier. These scattered data mainly result from the fusion cross section measurements by the evaporation residue detection method [34,42]. The reason for the underestimated values of cross sections in the above works have been discussed in details earlier [19-21, 30]. It is due to the difficulties in detecting the residues of low kinetic energy (which makes the underestimate of their yield and hence the fusion cross sections) particularly at low bombarding energies. That this is the fact has been shown by the accurate measurement of evaporation residues for the $^7$Li + $^{12}$C reaction in the reverse kinematics [22]. Thus if we exclude the low energy portion of the evaporation residue data (up to $E_{c.m.}/V_B \sim 3.5$) and consider ~15% uncertainty in the measured values of cross sections in general, it appears that all the systems show nearly identical reduced fusion cross sections.

Nevertheless the cross sections in the reduced form as shown above should be considered only in the spirit of systematic presentation of the data for a number of reactions together. Moreover the representation as shown above is not unique and there can be other forms of reduced cross sections as function of reduced energy. Finally, it is rather impossible to treat the reactions involving nuclei all throughout the periodic table using a single form of reduced cross section like the above.

# V. CONCLUSIONS

In this work we have measured the cross sections for the characteristic γ-rays of the residual nuclei following $^{6}$Li+$^{24}$Mg and $^{7}$Li+$^{24}$Mg reactions at energies ~2MeV below and more than three times above the Coulomb barrier.

From these γ-ray cross sections we determined the cross sections for different channels as well as the total fusion cross sections.

The Coupled Channel calculation (subject to the potential used) fail to reproduce the fusion cross sections at sub-barrier energies. The fusion cross sections are, however, found to be in good agreement with the total reaction cross sections obtained from the Optical Model calculations, at such energies. The decrease of the measured fusion cross sections at high bombarding energies is attributed to be mainly due to the quasi-elastic breakup reactions. We may thus conclude that the fusion cross sections measured for these two systems $^{6}$Li+$^{24}$Mg and $^{7}$Li+$^{24}$Mg do not show any kind of enhancement or inhibition in spite of the fact that the two projectiles are loosely bound nuclei.

Comparison with other Li-induced reactions reveals that these two systems behave in identical manner both in their fusion cross sections and angular momenta.

**ACKNOWLEDGEMENTS**

The authors would like to thank Mr. Sujib. Chatterjee and Mr. Pradipta. Kr. Das for their earnest support throughout the experiment especially in chamber designing and target preparation.

## REFERENCES

[1] A. Yoshida , C. Signorini, T. Fukuda, Y. Watanabe, N. Aoi, M. Hirai, M. Ishihara, H. Kobinata, Y. Mizoi, L. Mueller, Y. Nagashima, J. Nakano, T. Nomura,Y. H. Pu. and F. Scarlassara, Phys. Lett. B389, 457 (1996).

[2] K. E. Rehm, H. Esbensen, C. L. Jiang, B. B. Back, F. Borasi, B. Harss, R. V. F. Janssens, V. Nanal, J. Nolen, R. C. Pardo, M. Paul, P. Reiter, R. E. Segel, A. Sonzogni, J. Uusitalo, and A. H. Wuosmaa, Phys. Rev. Lett. 81, 3341 (1998).

[3] J. J. Kolata,V. Guimarães, D. Peterson, P. Santi, R. White-Stevens, P. A. De Young, G. F. Peaslee, B. Hughey, B. Atalla, M. Kern, P. L. Jolivette, J. A. Zimmerman, M. Y. Lee, F. D. Becchetti, E. F. Aguilera, E. Martinez-Quiroz, and J. D. Hinnefeld, Phys. Rev. Lett. 81, 4580 (1998).

[4] E. A. Benjamin, A. Lépine-Szily, D. R. Mendes Junior, R. Lichtenthäler, V. Guimarães, P. R. S. Gomes, L. C. Chamon, M. S. Hussein, A. M. Moro, A. Arazi, I. Padron, J. Alcantara Nuñez, M. Assuncão, A. Barioni, O. Camargo Jr., R. Z. Denke, P. N. de Faria and K. C. C. Pires, Phys. Lett. B647, 30 (2007).

[5] P. R. S. Gomes, I. Padron, E. Crema, O. A. Capurro, J. O. Fernández Niello, G. V. Martí, A. Arazi, M. Trotta, J. Lubian, M. E. Ortega, A. J. Pacheco, M. D. Rodriguez, J. E. Testoni, R. M. Anjos, L. C. Chamon, M. Dasgupta, D. J. Hinde and K. Hagino, Phys. Lett B 634, 356 (2006).

[6] M. Dasgupta, D. J. Hinde, R. D. Butt, R. M. Anjos, A. C.Berriman, N. Carlin, P. R. S. Gomes, C. R. Morton, J. O. Newton, A. Szanto de Toledo, and K. Hagino, Phys. Rev. Lett. 82, 1395 (1999).

[7] C. Signorini, Z. H Liu, Z. C. Li, K. E. G. Löbner, L. $M\ddot{u}llar$, M. Ruan, K. Rudolph, F. Soramel, C. Zotti, A. Andrighetto, L. Stroe, A. Vitturi, H. Q. Zhang, Eur. Phys. Jour. A 5, 7 (1999).

[8] M. Dasgupta, D. J. Hinde, K. Hagino, S. B. Moraes, P. R. S. Gomes, R. M Anjos, R. D. Butt, A. C. Berriman, N. Carlin, C. R. Morton, J. O. Newton, and A. Szantode Toledo, Phys. Rev. C66, 041602(R) (2002).

[9] V. Tripathi, A. Navin, K. Mahata, K. Ramachandran, A. Chatterjee and S. Kailas, Phys. Rev. Lett.88, 172701 (2002).

[10] A. Mukherjee, Subinit Roy, M. K. Pradhan, M. Saha. Sarkar, P. Basu, B. Dasmahapatra, T. Bhattacharya, S. K. Basu, A. Chatterjee, V. Tripathi, S. Kailas, Phys. Lett. B636, 91, (2006).

[11] S. B Moraes, P. R. S. Gomes, J. Lubian, J. J. S. Alves, R. M Anjos, M. M. Sant´Anna, I. Padron, C. Muri, R. Liguori Neto, and N. Added., Phys. Rev. C61, 064608 (2000).

[12] P. R. S. Gomes, I. Padron, M. D. Rodríguez, G. V. Martí, R. M. Anjos, J. Lubian, R. Veiga, R. Liguori Neto, E. Crema, N. Added, L. C. Chamon, J. O. Fernández Niello, O. A. Capurro, A. J. Pacheco, J. E. Testoni, D. Abriola, A. Arazi, M. Ramírez, M. S. Hussein Phys. Lett. B601, 20-26 (2004).

[13] C. Beck, F. A. Souza, N. Rowley, S. J. Sanders, N. Aissaoui, E. E. Alonso, P. Bednarczyk, N. Carlin, S. Courtin, A. Diaz-Torres, A. Dummer, F. Haas, A. Hachem, K. Hagino, F. Hoellinger, R. V. F. Janssens, N. Kintz, R. Liguori Neto, E. Martin, M. M. Moura, M. G. Munhoz, P. Papka, M. Rousseau, A. Sànchezi Zafra, O. Stézowski, A. A. suaide, E. M. Szanto, A. Szanto de Toledo, S. Szilner and J. Takahashi, Phys. Rev. C67, 054602 (2003).

[14] R. M. Anjos, C. Muri, J. Lubian, P. R. S. Gomes, I. Padron, J. J. S. Alves, G. V. Martí, J. O. Fernández Niello, A. J. Pacheco, O. A. Capurro, D. Abriola, J. E. Testoni, M. Ramirez, R. Liguori Neto, and N. Added, Phys. Lett. B534, 45 (2002).

[15] I. Padron, P. R. S. Gomes, R. M. Anjos, J. Lubian, C. Muri, J. J. S. Alves, G. V. Martí, M. Ramírez, A. J. Pacheco, O. A. Capurro, J. O. Fernández Niello, J. E. Testoni, D. Abriola, and M. R. Spinella Phys. Rev. C66, 044608 (2002).

[16] G. V. Martí, P. R. S. Gomes, M. D. Rodríguez, J. O. Fernández Niello, O. A. Capurro, A. J. Pacheco, J. E. Testoni, M. Ramírez, A. Arazi, O. A. Capurro, I. Padron, R. M. Anjos, J. Lubian, and E. Crema, Phys. Rev. C71, 027602 (2005).

[17] K. Kalita, S. Verma, R. Singh, J. J. Das, A. Jhingan, N. Madhavan, S. Nath, T.Varughese, P. Sugathan, V. V. Parkar, K. Mahata, K. Ramachandran, A. Shrivastava, A. Chatterjee, S. Kailas, S. Barua, P. Basu, H. Majumdar, M. Sinha, R. Bhattacharya, and A. K. Sinha, Phys. Rev. C73, 024609 (2006).

[18] A. Mukherjee, U. Datta Pramanik, S. Chattopadhyay, M. Saha Sarkar, A. Goswami, P. Basu, S. Bhattacharya, M. L. Chatterjee, and B. Dasmahapatra, Nucl. Phys. A645, 13 (1999).

[19] A. Mukherjee and B. Dasmahapatra, Phys. Rev. C63, 017604 (2000).

[20] A. Mukherjee, U. Datta Pramanik, M. Saha Sarkar, A. Goswami, P. Basu, S. Bhattacharya, S. Sen, M. L. Chatterjee, and B. Dasmahapatra, Nucl. Phys. A596, 299 (1996).

[21] A. Mukherjee, U. Datta Pramanik, S. Chattopadhyay, M. Saha Sarkar, A. Goswami, P. Basu, S. Bhattacharya, M. L. Chatterjee, and B. Dasmahapatra, Nucl. Phys. A635, 305 (1998).

[22] A. Mukherjee, M. Dasgupta, D. J. Hinde, H. Timmers, R. D. Butt and P. R. S. Gomes Phys. Lett. B526, 295 (2002).

[23] A. Mukherjee and B. Dasmahapatra, Nucl. Phys. A 614, 238 (1997).

[24] C.J.S. Scholz, L. Ricken and E. Kuhlmann, Z. Phys. A 325, 203 (1986).

[25] R. A. Dayras, R. G. Stokstad, Z. E. Switkowski and R. M. Wieland, Nucl. Phys. A265, 153 (1976).

[26] Y. D. Chan, H. Bohn, R. Vandenbosch, K. G. Bernhardt, J. G. Cramer, R. Sielemann and L. Green; Nucl. Phys. A 303, 500 (1978).

[27] J. J. Kolata, R. M. Freeman, F. Hass, B. Heusch and A. Gallmann, Phys. Rev. C19, 408 (1979).

[28] J. J. Kolata, R. M. Freeman, F. Hass, B. Heusch and A. Gallmann, Phys. Rev. C19, 2237 (1979).

[29] F. Pühlhofer, Nucl. Phys. A 280, 267 (1977). Program CASCADE: Rapport GSI, Darmstadt (1976), unpublished.

[30] M Ray, A. Mukherjee, M. Saha. Sarkar, A. Goswami, S. Roy, S. Saha, R. Bhattacharya, B. R. Behara, S. K. Datta, and B. Dasmahapatra Phys. Rev. C68, 067601 (2003).

[31] J. E. Poling, E. Norbeck and R. R.Carlson, Phys. Rev. C13, 648 (1976).

[32] A. Mukherjee PhD. Thesis, CU, 1998, unpublished.

[33] S. L. Tabor. L. C. Dennis and K. Abdo Nucl. Phys. A 391, 458 (1982).

[34] L. C. Dennis, K. M. Abdo, A. D. Frawley, and K. W. Kemper, Phys. Rev. C26, 981 (1982).

[35] Mandira Sinha, H. Majumdar, R. Bhattacharya, P. Basu. Subinit Roy, M. Biswas, R. Palit, I. Mazumdar, P. K. Joshi, H. C. Jain, and S. Kailas Phys. Rev. C76, 027603 (2007).

[36] R. M. Anjos, V. Guimarães, N. Added, N. Carlin Filho, M. M. Coimbra, L. Fante, Jr., M. C. S. Figueira, E. M. Szanto, C. F. Tenreiro, and A. Szanto de Toledo, Phys. Rev. C42, 354 (1990).

[37] K. Hagino, N. Rowley, A. T. Kruppa, Comput. Phys. Commun. 123, 143 (1999).

[38] N. Keeley, K. W. Kemper and K. Rusek, Phys. Rev. C65, 014601 (2001).

[39] R. G. Stokstad, Z. E. Switkowski, R. A. Dayras and R. M. Wieland, Phys. Rev. Lett. 37, 888 (1976).

[40] Q. Haider and B. Čujec Nucl. Phys. A429, 116 (1984).

[41] Louis. C. Vaz, John. M. Alexander, and G. R. Satchler, Phys. Rep. 69, 373 (1981).

[42] J. F. Mateja, J. Garman, D. E. Fields, R. L. Kozub, A. D. Frawley and L. C. Dennis, Phys. Rev. C30, 134 (1984).

[43] F. Ajzenberg-Selove, Nucl. Phys. A 506, 1 (1990) ; A 523, 1 (1991) ; A 460, 1 (1986); A 475, 1 (1987).

TableI: Identification of contaminant peaks in the $^{6,7}Li+^{24}Mg$ γ-ray spectra (Fig.3).

| Label | Energy(MeV) | Transition | Origin |
|---|---|---|---|
| A | 0.136 | $^{181}$Ta (0.136 → 0) | $^{181}$Ta(n n′) |
| B | 0.166 | $^{181}$Ta (0.302 → 0.136) | $^{181}$Ta(n n′) |
| C | 0.239 | Th-series | radioactivity |
| D | 0.296 | Ra-series | radioactivity |
| E | 0.302 | $^{181}$Ta (0.302 → 0) | $^{181}$Ta(n n′) |
| F | 0.352 | Ra-series | radioactivity |
| G | 0.583 | Th-series | radioactivity |
| H | 0.596<br>0.609 | $^{74}$Ge (0.596 → 0)<br>Ra-series | $^{74}$Ge(n n′)<br>radioactivity |
| I | 0.691 | $^{72}$Ge (0.691 → 0) | $^{72}$Ge(n n′) |
| J | 0.844<br>0.847 | $^{27}$Al (0.844 → 0)<br>$^{56}$Fe (0.847 → 0) | $^{27}$Al(n n′)<br>$^{56}$Fe(n n′) |
| K | 0.909 | Th-series | radioactivity |
| L | 0.967 | Th-series | radioactivity |
| M | 1.238<br>1.238 | $^{56}$Fe (2.085 → 0.847)<br>Ra-series | $^{56}$Fe(n n′)<br>radioactivity |
| N | 1.434 | $^{52}$Cr (1.434 → 0) | $^{52}$Cr(n n′) |
| O | 1.461 | $^{40}$Ar (1.461 → 0) | radioactivity |
| P | 2.615 | Th-series | radioactivity |

## FIGURE CAPTIONS

Fig.1: Energy level diagram for the $^{6}Li + ^{24}Mg$ reaction. The numbers attached to the ground states give the Q-values of the respective channels. The energy region investigated in the present work is cross-hatched. The transitions shown are those for which the γ-ray cross sections were measured. The highest levels indicated are particle unstable.

Fig.2: Energy level diagram for the $^{7}Li + ^{24}Mg$ reaction. For other details see caption of Fig.1.

Fig.3: (a) - (c) : Gamma-ray spectra obtained for $^{6}Li + ^{24}Mg$ (top spectrum which is displaced vertically by the indicated factor) and $^{7}Li + ^{24}Mg$ (bottom spectrum) at $E_{lab}$ = 11 MeV, obtained with target of natural Mg. The contaminant lines are marked by alphabets and are identified in TableI. The γ-rays belonging to $^{6,7}Li + ^{24}Mg$ reactions are indicated by showing them within the square brackets. The γ-ray lines arising due to the reactions with $^{25}Mg$ and $^{26}Mg$ present in the natural $^{24}Mg$ target have been identified by the symbol ($), (#) respectively.

Fig.4: Theoretical branching factors $f_\gamma$ (= $\sigma_\gamma$ / $\sigma_{ch}$ ) for the decay of the residual nuclei following compound nucleus formation, calculated with the code CASCADE and the γ-ray branching factors from Ref. [43].

Fig.5: Cross sections for different exit channels of the reaction $^{6}Li + ^{24}Mg$ using the experimental γ -ray cross sections and the $f_\gamma$ values shown in fig 4(a). The open circles (○) and open triangles (△) show data obtained at IOP. The solid circles (●) and solid triangles (▲) show data obtained at TIFR. The error bars show the absolute total error. The solid lines are the calculated channel cross sections. For clarity the data have been displaced vertically by the indicated factors. The cross sections for the pn + $^{28}Si$ channel measured from 1.779 MeV and 2.837 MeV γ-rays of $^{28}Si$ obtained are shown with the symbols triangles and circles respectively. The cross sections for the pα + $^{25}Mg$ channel measured from 0.585 and from the sum of 0.390 and 0.975MeV γ-rays of $^{25}Mg$ obtained are shown with the symbols triangles and circles respectively.

Fig.6: Cross sections for different exit channels of the reaction $^{7}Li + ^{24}Mg$ using the experimental γ -ray cross sections and the $f_\gamma$ values shown in fig 4(b). The open circles (○), triangles (△) and squares (□) show data obtained at IOP. The solid circles (●), triangles (▲) and squares (■) show data obtained at TIFR. The error bars show the absolute total error. The solid lines are the calculated channel cross sections. For clarity the data have been displaced vertically by the indicated factors. The cross sections for the pn + $^{29}Si$ channel measured from 1.273 MeV, 2.028 MeV and 1.596MeV γ-rays of $^{29}Si$ obtained are shown with the symbols triangles, circles and squares respectively.

Fig.7: a) Theoretical branching factors $F_\gamma$ for the decay of the residual nuclei (following compound nucleus formation) formed in the reaction $^6$Li + $^{24}$Mg. The $F_\gamma$ values were calculated using the sum of the theoretical γ-ray cross sections as mentioned in the plots. The calculations were done with the code CASCADE. $F_\gamma$ calculated with 0.390, 0.975, 1.779, 2.837, 0.451 and 0.891 MeV γ-rays are shown by the curve A. B represents the same including 1.368 and 1.014 MeV γ-rays. b) Fusion cross sections for the reaction $^6$Li + $^{24}$Mg using different sets of $F_\gamma$ values as mentioned in the figure shown above. The error bars show the absolute total error. The solid line represents the total reaction cross sections calculated using optical model [31]. For details see text.

Fig.8: a) Theoretical branching factors $F_\gamma$ for the decay of the residual nuclei (following compound nucleus formation) formed in the reaction $^7$Li + $^{24}$Mg. The $F_\gamma$ values were calculated using the sum of the theoretical γ-ray cross sections as mentioned in the plots. The calculations were done with the code CASCADE. $F_\gamma$ calculated with 2.028, 1.596,1.808, 0.440 and 0.417 MeV γ-rays are shown by the curve A. B represents the same including 1.779, 2.837, 0.390 and 0.975 MeV γ-rays. b) Fusion cross sections for the reaction $^7$Li + $^{24}$Mg using different sets of $F_\gamma$ values as mentioned in the figure shown above. The error bars show the absolute total error. The solid line represents the total reaction cross sections calculated using optical model [31]. For details see text.

Fig.9: Critical angular momenta ($l_{cr}$) and grazing angular momentum ($l_{gr}$) as a function of excitation energy of the compound nucleus formed by the different incident channels are shown with solid circles (●) and the solid line respectively. $l_{cr}$ are obtained from experimental fusion cross sections and $l_{gr}$ are obtained from the optical model calculations using the parameters from fitting the elastic scattering data (see text for details). Horizontal dashed lines represent the compound nucleus excitation energy corresponding to $E_{c.m.} = B_C$ [where $B_C = \frac{Z_P Z_T e^2}{1.70(A_P^{1/3} + A_T^{1/3})}$], for each system.

Fig.10: Fusion cross sections for the $^{6,7}$Li + $^{24}$Mg reaction compared with the theoretical model calculations. The solid circles (●) show the present measurements. The error bars show the absolute error. The solid and dotted lines represent the total reaction cross sections calculated using optical model [31] and fusion cross sections by CCFULL calculations [37] respectively. For details see text. The parameters of the potential used in the above calculations are as follows:
OM : $V_o$ = 172MeV, $r_0$ = 1.4fm, a = 0.73fm.
CCFULL (uncoupled) : $V_o$ = 130MeV, $r_0$ = 0.97fm, a = 0.63fm.

Fig.11: Total fusion cross sections (with error) for Li-induced reactions: $^{6,7}$Li + $^{24}$Mg, $^{6,7}$Li + $^{27}$Al and $^{6,7}$Li + $^{28}$Si. The arrow indicates the position of Coulomb barrier energy $B_C$ for the $^6$Li + $^{24}$Mg system.

[where, $B_C = \frac{Z_P\, Z_T\, e^2}{1.70(A_P^{1/3} + A_T^{1/3})}$] . The solid line represents the total reaction cross sections for the $^6$Li + $^{24}$Mg reaction calculated using optical model. For details see text.

Fig.12: Reduced fusion excitation functions for the $^6$Li induced reactions on light mass targets. The barrier parameters $V_B$ and $R_B$ were obtained from the systematics proposed by Vaz et. al. [Ref. 41]. The data for different reactions are marked by the following symbols : ● $^6$Li + $^{24}$Mg Present Work; ▽ $^6$Li + $^{27}$Al Padron et al.(2002) [15]; △ $^6$Li + $^{12}$C Mukherjee et al.(1996) [20]; ◇ $^6$Li + $^{13}$C Mukherjee et al.(1998) [21]; × $^6$Li + $^{16}$O Mukherjee et al.(1999), Scholz et al.(1986) [18, 24]; + $^6$Li + $^{16}$O Mateja et al.(1984) [42]; ▲ $^6$Li + $^{12}$C Dennis et al.(1982) [34]; ◆ $^6$Li + $^{13}$C Dennis et al.(1982) [34].

Fig.13: Reduced fusion excitation functions for the $^7$Li induced reactions on light mass targets. The barrier parameters $V_B$ and $R_B$ were obtained from the systematics proposed by Vaz et. al. [Ref. 41]. The data for different reactions are marked by the following symbols : ● $^7$Li + $^{24}$Mg Present Work; □ Mray et al.(2003) [30]; × $^7$Li + $^{16}$O Mukherjee et al.(1999), Scholz et al.(1986) [18, 24]; + $^7$Li + $^{16}$O Mateja et al.(1984) [42]; ▽ $^7$Li + $^{27}$Al Padron et al.(2002), Kalita et al.(2006) [15,17]; △ $^7$Li + $^{12}$C Mukherjee et al.(1996) [20]; ◇ $^7$Li + $^{13}$C Mukherjee et al.(1998) [21]; ○ $^7$Li + $^{28}$Si Sinha et al.(2007) [35]; ▲ $^7$Li + $^{12}$C Dennis et al.(1982) [34]; ◆ $^7$Li + $^{13}$C Dennis et al.(1982) [34].

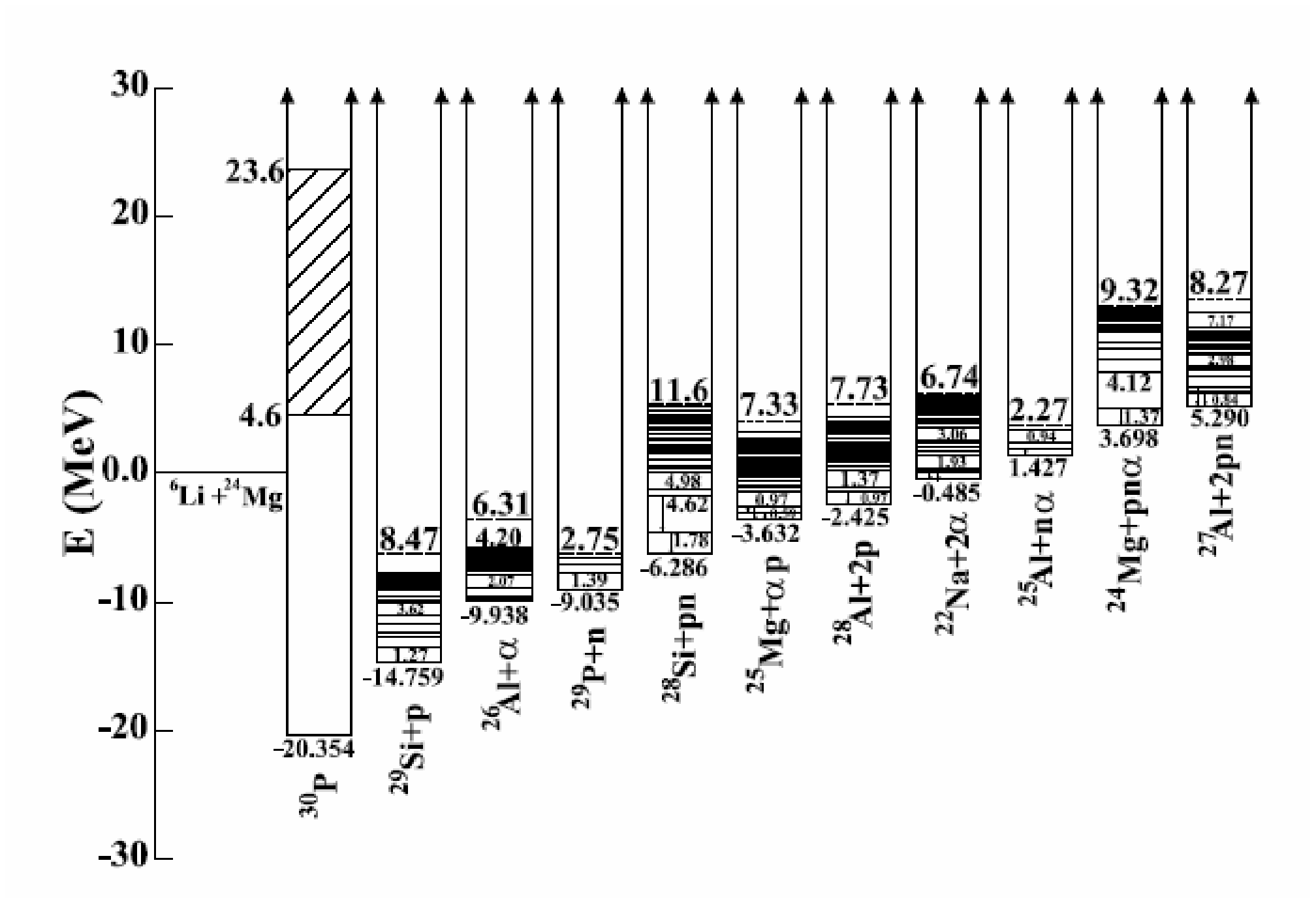

Fig1.

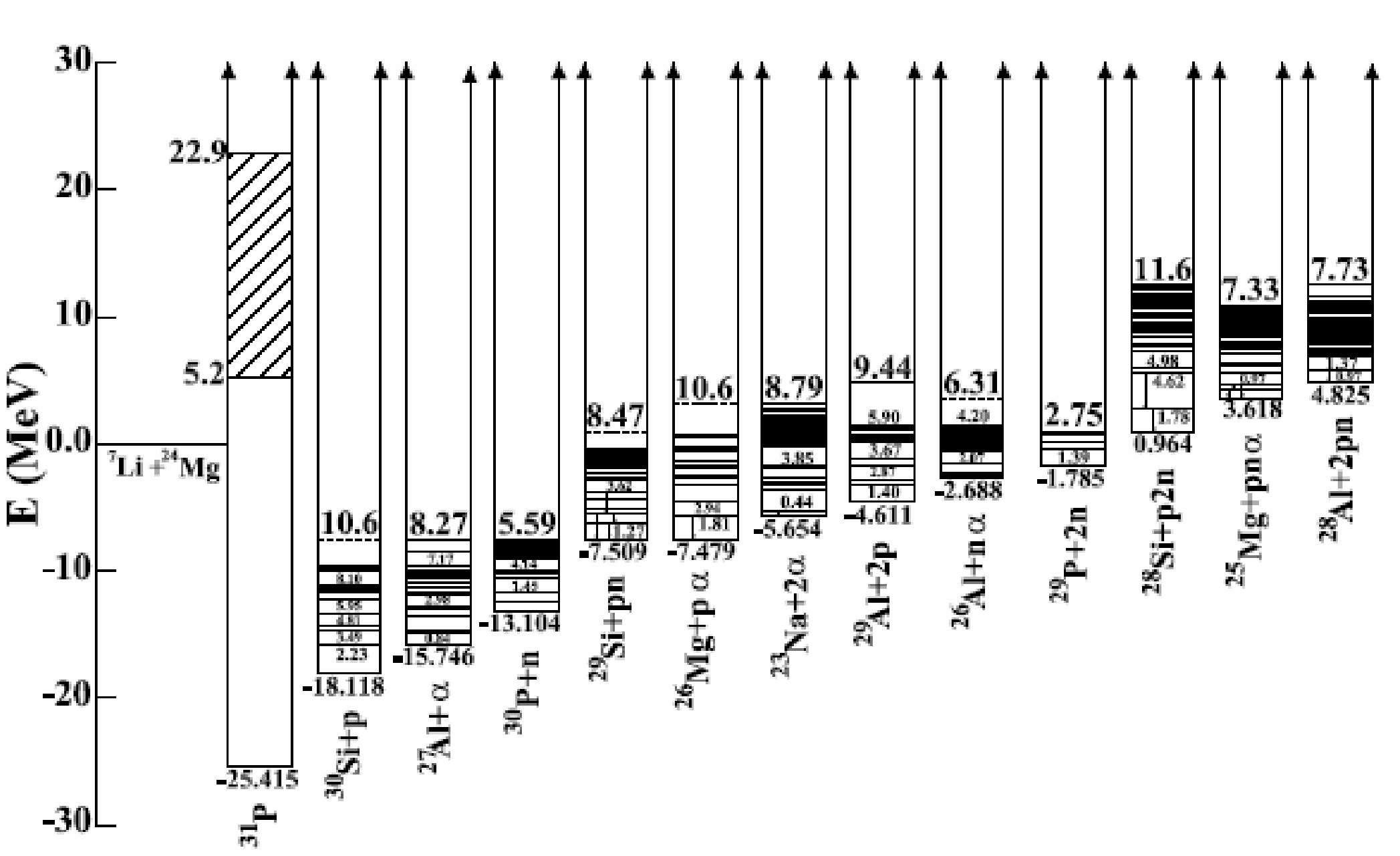

Fig2.

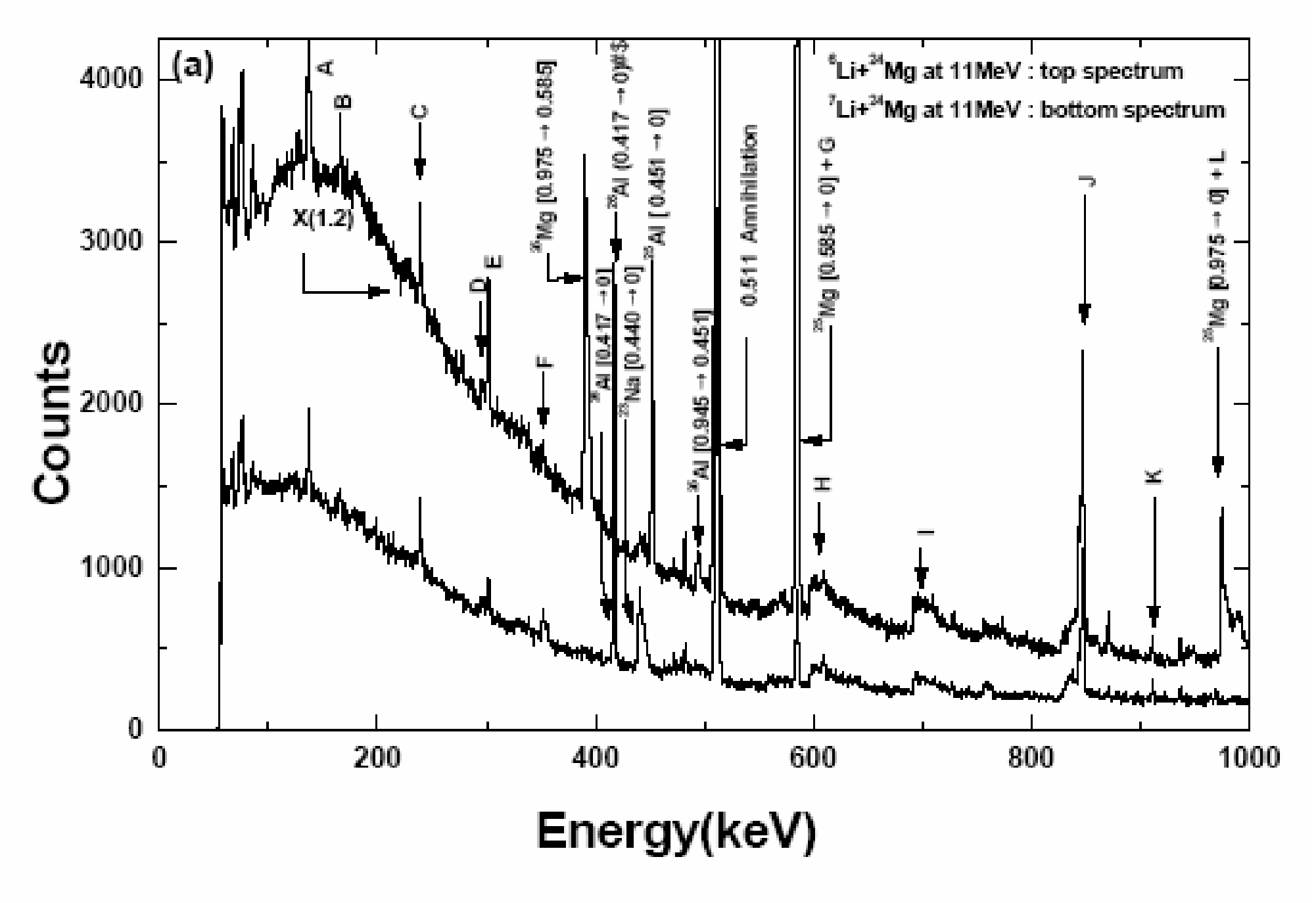
(a)
6Li+24Mg at 11MeV : top spectrum
7Li+24Mg at 11MeV : bottom spectrum
Counts
Energy(keV)
0
1000
2000
3000
4000
0
200
400
600
800
1000
A
B
C
D
E
F
H
I
J
K
X(1.2)
Mg [0.975 → 0.585]
Al [0.417 → 0]
Na [0.440 → 0]
Al [ 0.451 → 0]
Al [0.945 → 0.451]
0.511 Annihilation
Mg [0.585 → 0] + G
Mg [0.975 → 0] + L

Fig3(a).

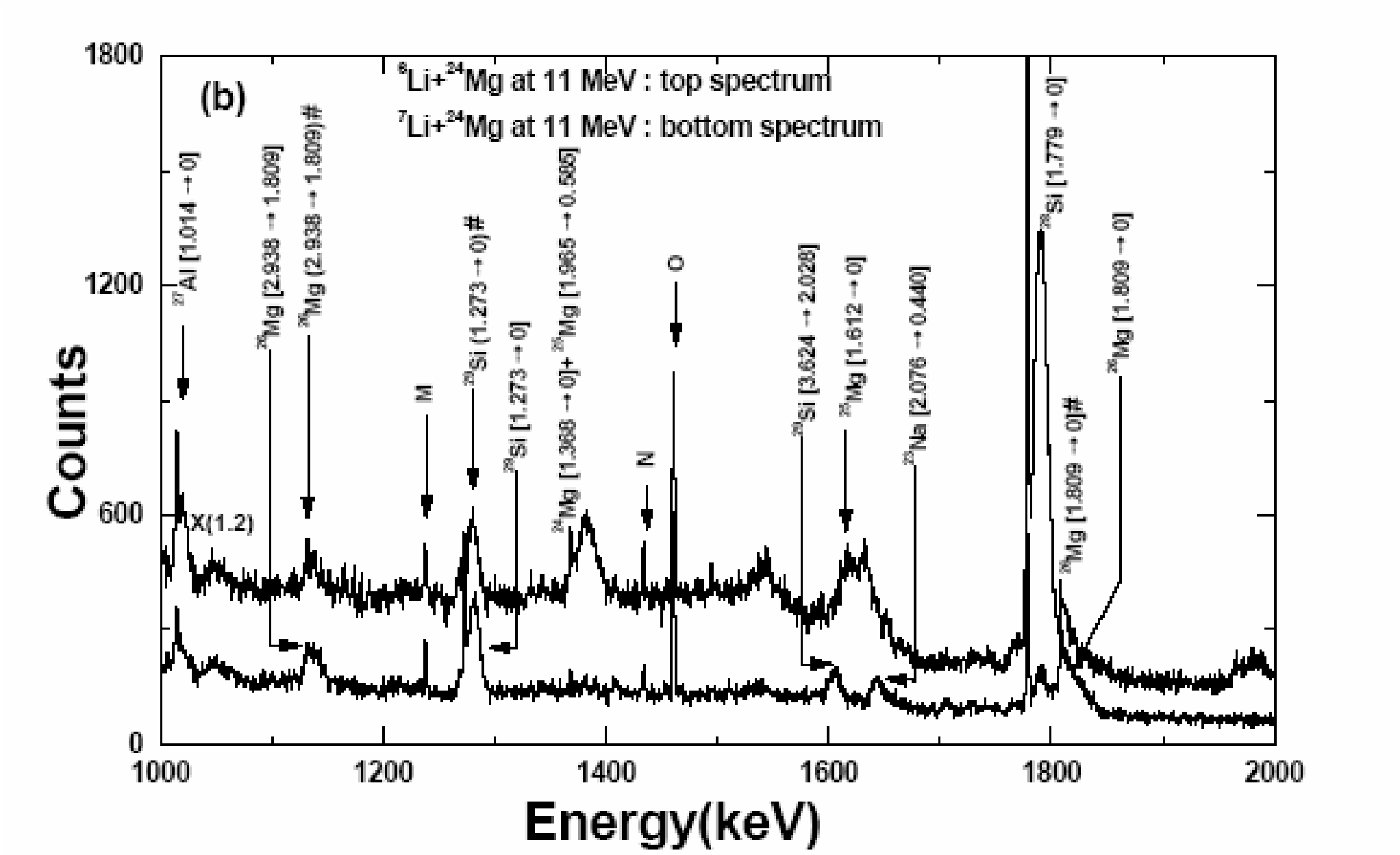
(b)
6Li+24Mg at 11 MeV : top spectrum
7Li+24Mg at 11 MeV : bottom spectrum
27Al [1.014 → 0]
X(1.2)
26Mg [2.938 → 1.809]
26Mg [2.938 → 1.809]#
M
29Si [1.273 → 0]#
29Si [1.273 → 0]
24Mg [1.368 → 0]+ 25Mg [1.965 → 0.585]
N
O
29Si [3.624 → 2.028]
25Mg [1.612 → 0]
23Na [2.076 → 0.440]
28Si [1.779 → 0]
26Mg [1.809 → 0]#
26Mg [1.809 → 0]
Counts
1800
1200
600
0
1000
1200
1400
1600
1800
2000
Energy(keV)

Fig3(b).

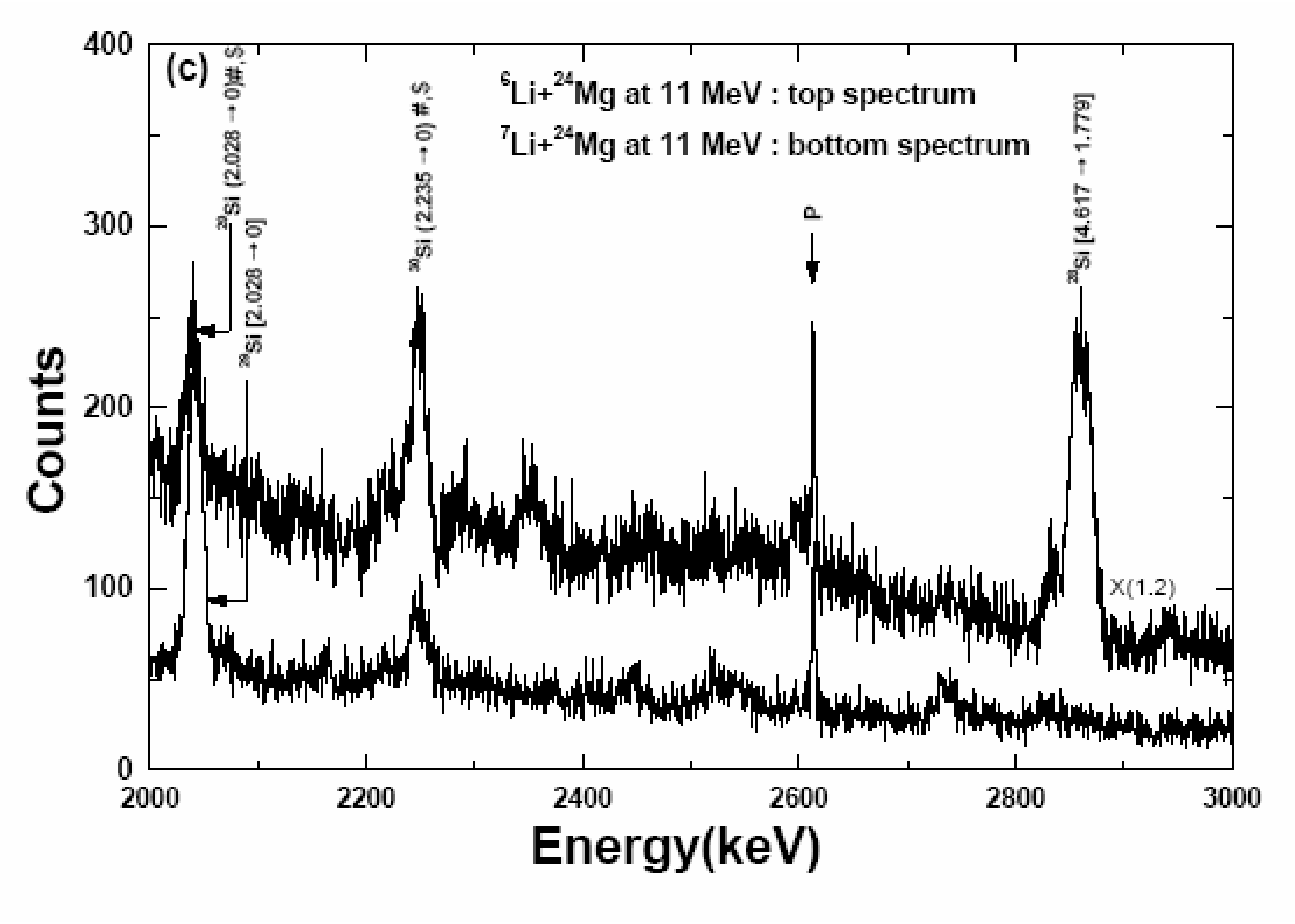
(c)
$^{6}$Li+$^{24}$Mg at 11 MeV : top spectrum
$^{7}$Li+$^{24}$Mg at 11 MeV : bottom spectrum
$^{29}$Si (2.028 → 0)#,S
$^{29}$Si [2.028 → 0]
$^{30}$Si (2.235 → 0) #,S
P
$^{28}$Si [4.617 → 1.779]
X(1.2)
Counts
Energy(keV)

Fig3(c).

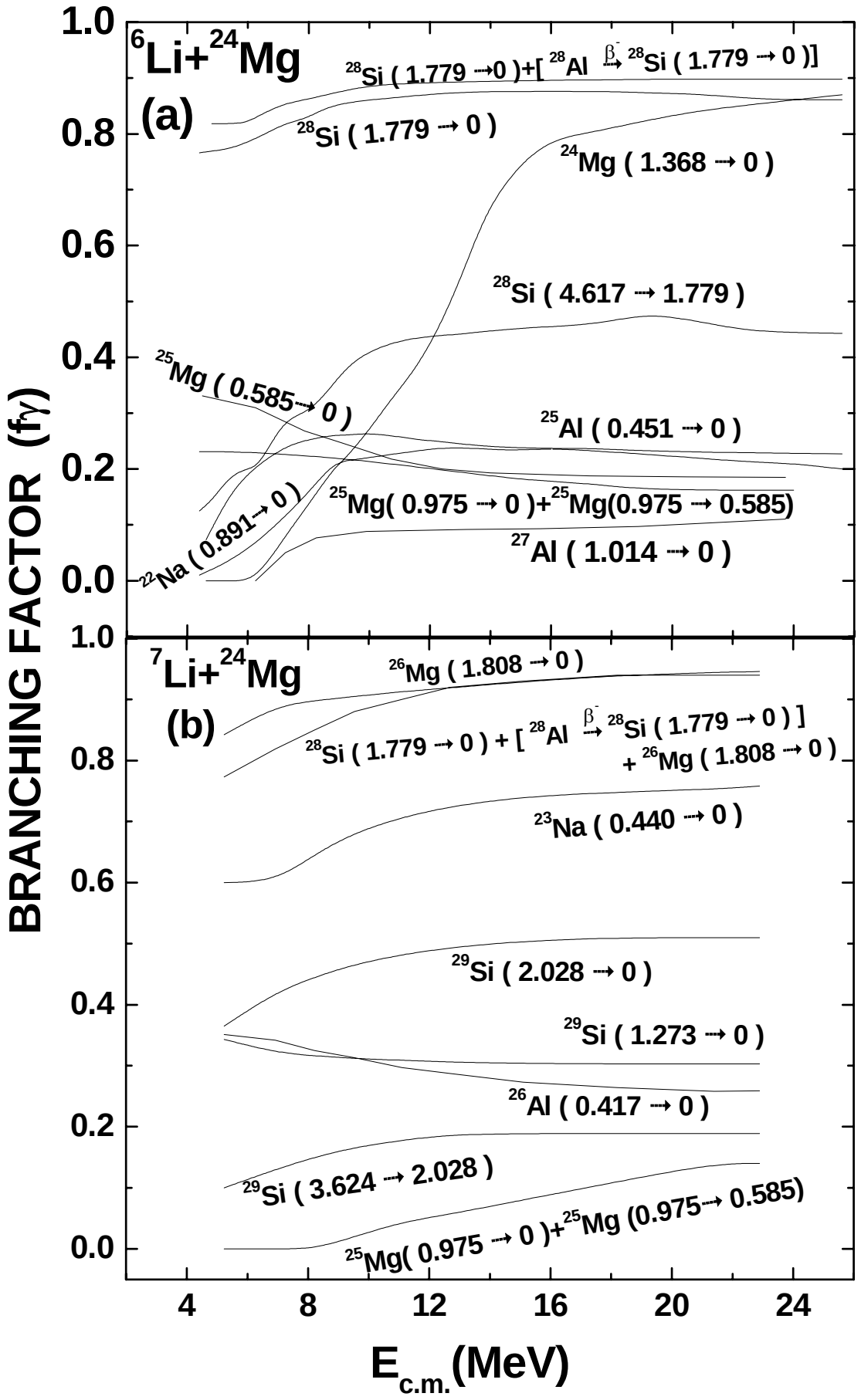

BRANCHING FACTOR (fγ)
6Li+24Mg
(a)
28Si ( 1.779 →0 )+[ 28Al β- → 28Si ( 1.779 → 0 )]
28Si ( 1.779 → 0 )
24Mg ( 1.368 → 0 )
28Si ( 4.617 → 1.779 )
25Mg ( 0.585→ 0 )
25Al ( 0.451 → 0 )
22Na ( 0.891→ 0 )
25Mg( 0.975 → 0 )+25Mg(0.975 → 0.585)
27Al ( 1.014 → 0 )
7Li+24Mg
(b)
26Mg ( 1.808 → 0 )
28Si ( 1.779 → 0 ) + [ 28Al β- → 28Si ( 1.779 → 0 ) ]
+ 26Mg ( 1.808 → 0 )
23Na ( 0.440 → 0 )
29Si ( 2.028 → 0 )
29Si ( 1.273 → 0 )
26Al ( 0.417 → 0 )
29Si ( 3.624 → 2.028 )
25Mg( 0.975 → 0 )+25Mg (0.975→ 0.585)
Ec.m.(MeV)

Fig4.

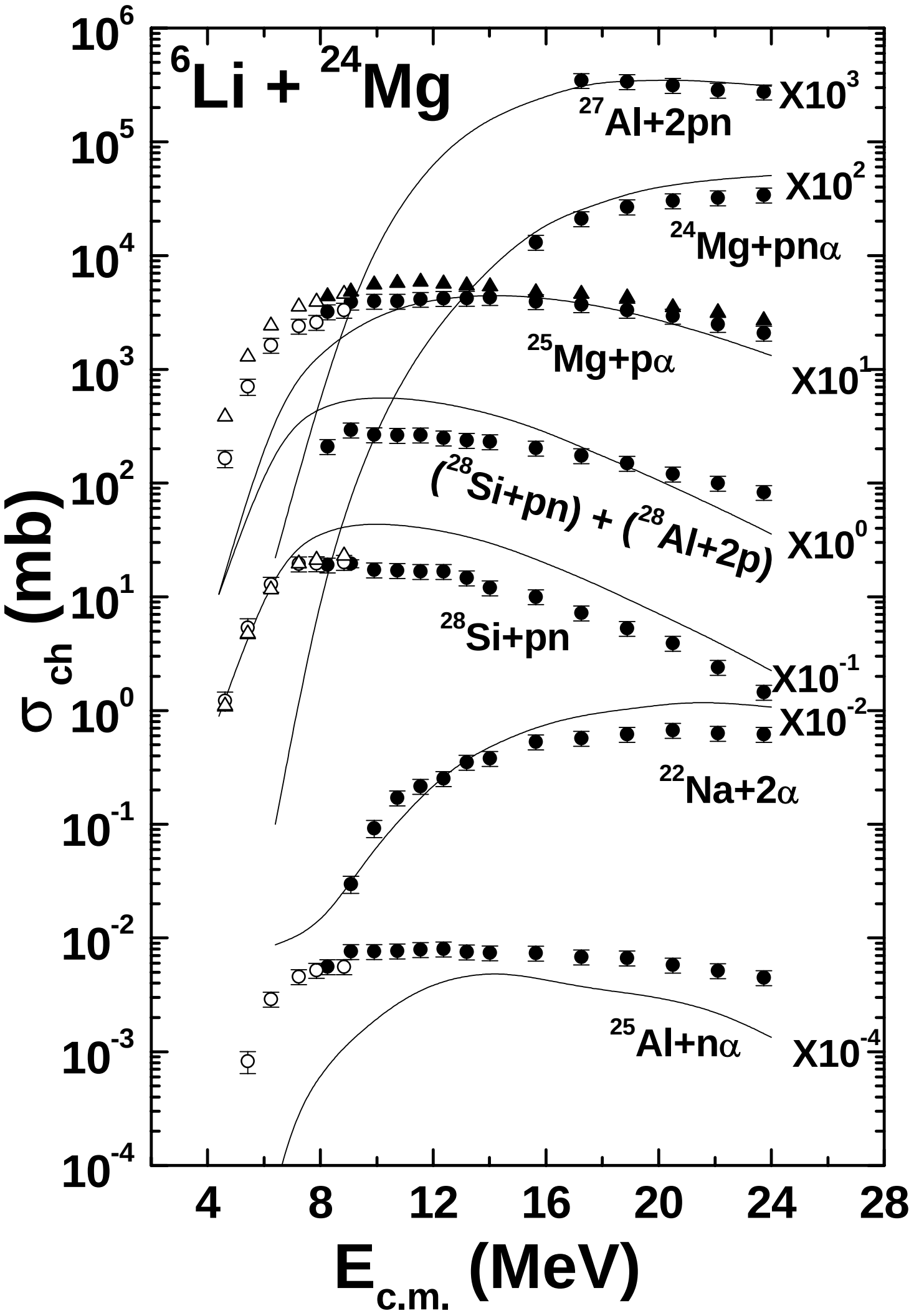

Fig5.

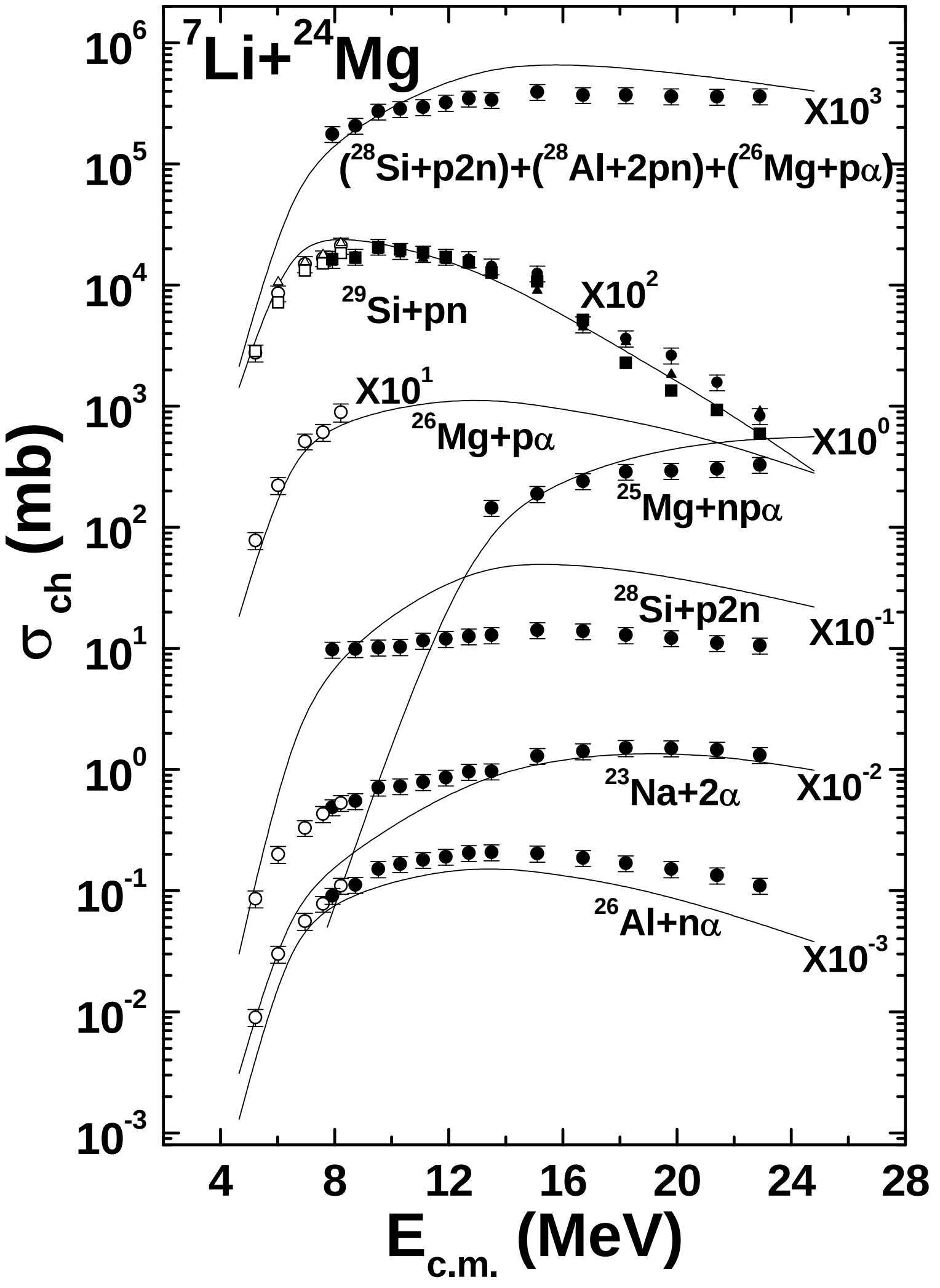

Fig6.

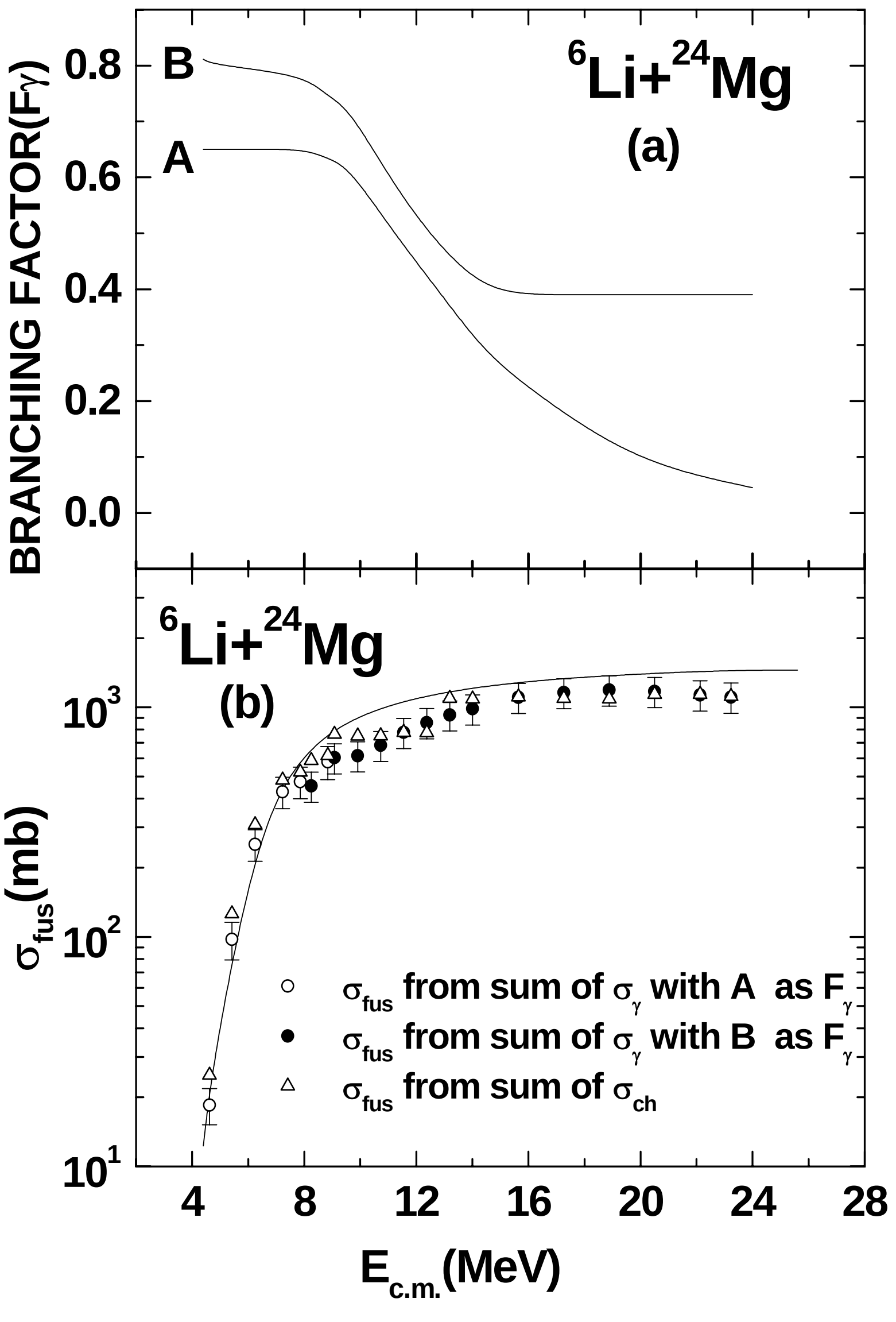

BRANCHING FACTOR(Fγ)
0.8
0.6
0.4
0.2
0.0
B
A
6Li+24Mg
(a)
σfus(mb)
10^3
10^2
10^1
6Li+24Mg
(b)
σfus from sum of σγ with A as Fγ
σfus from sum of σγ with B as Fγ
σfus from sum of σch
4
8
12
16
20
24
28
Ec.m.(MeV)

Fig7.

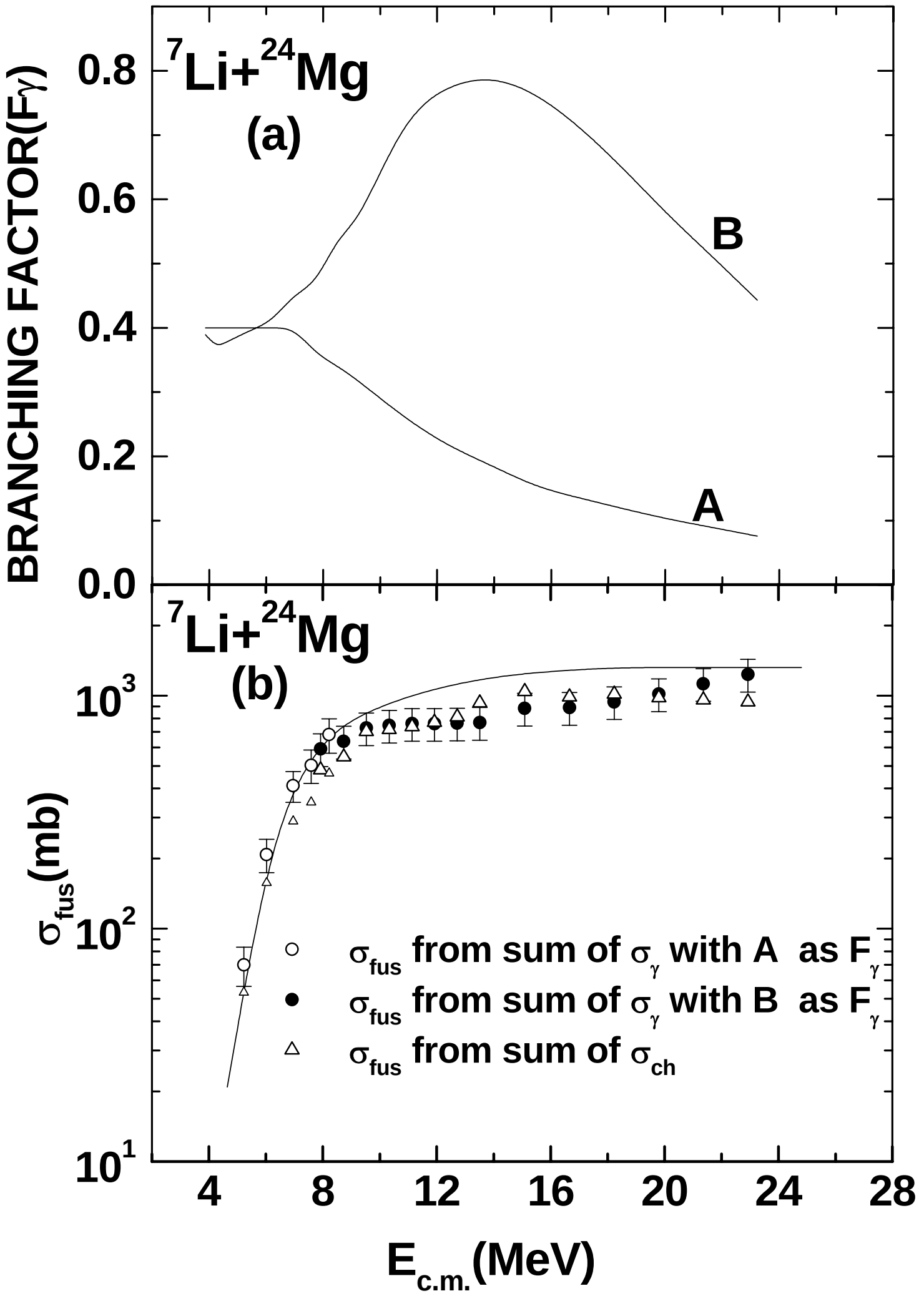

7Li+24Mg
(a)
BRANCHING FACTOR(Fγ)
0.0
0.2
0.4
0.6
0.8
B
A
7Li+24Mg
(b)
σfus(mb)
10¹
10²
10³
σfus from sum of σγ with A as Fγ
σfus from sum of σγ with B as Fγ
σfus from sum of σch
4
8
12
16
20
24
28
Ec.m.(MeV)

Fig8.

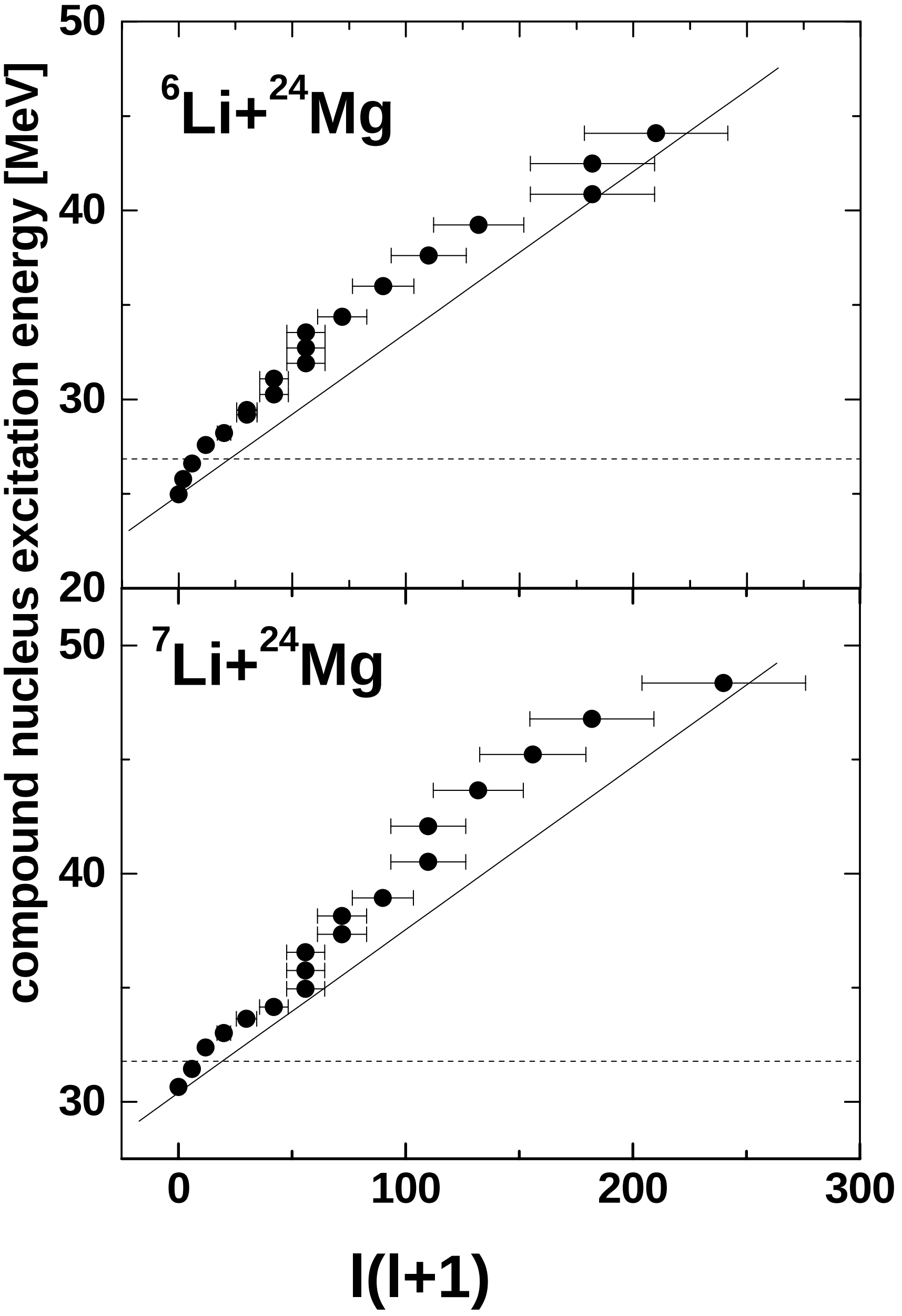

Fig9.

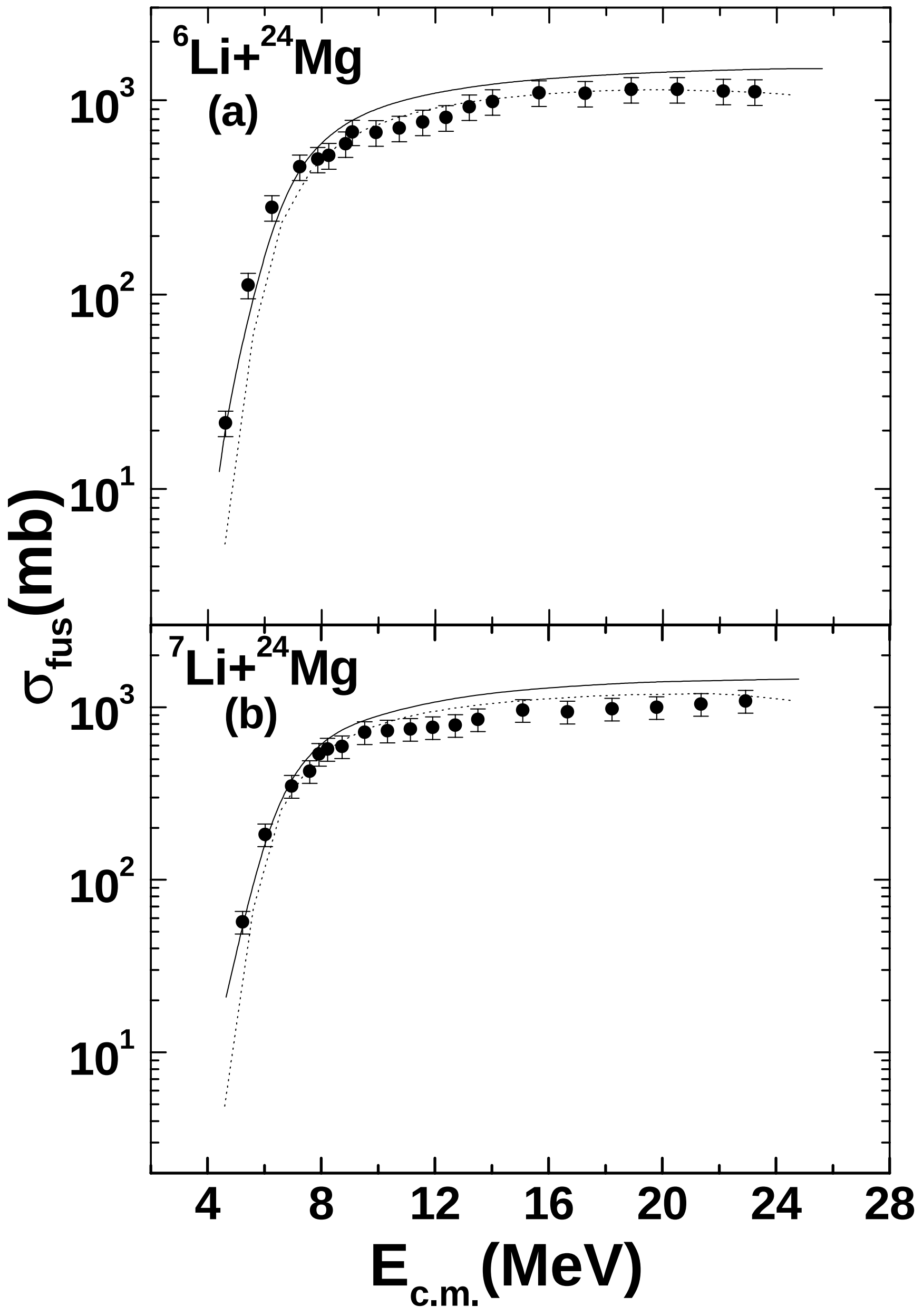

Fig10.

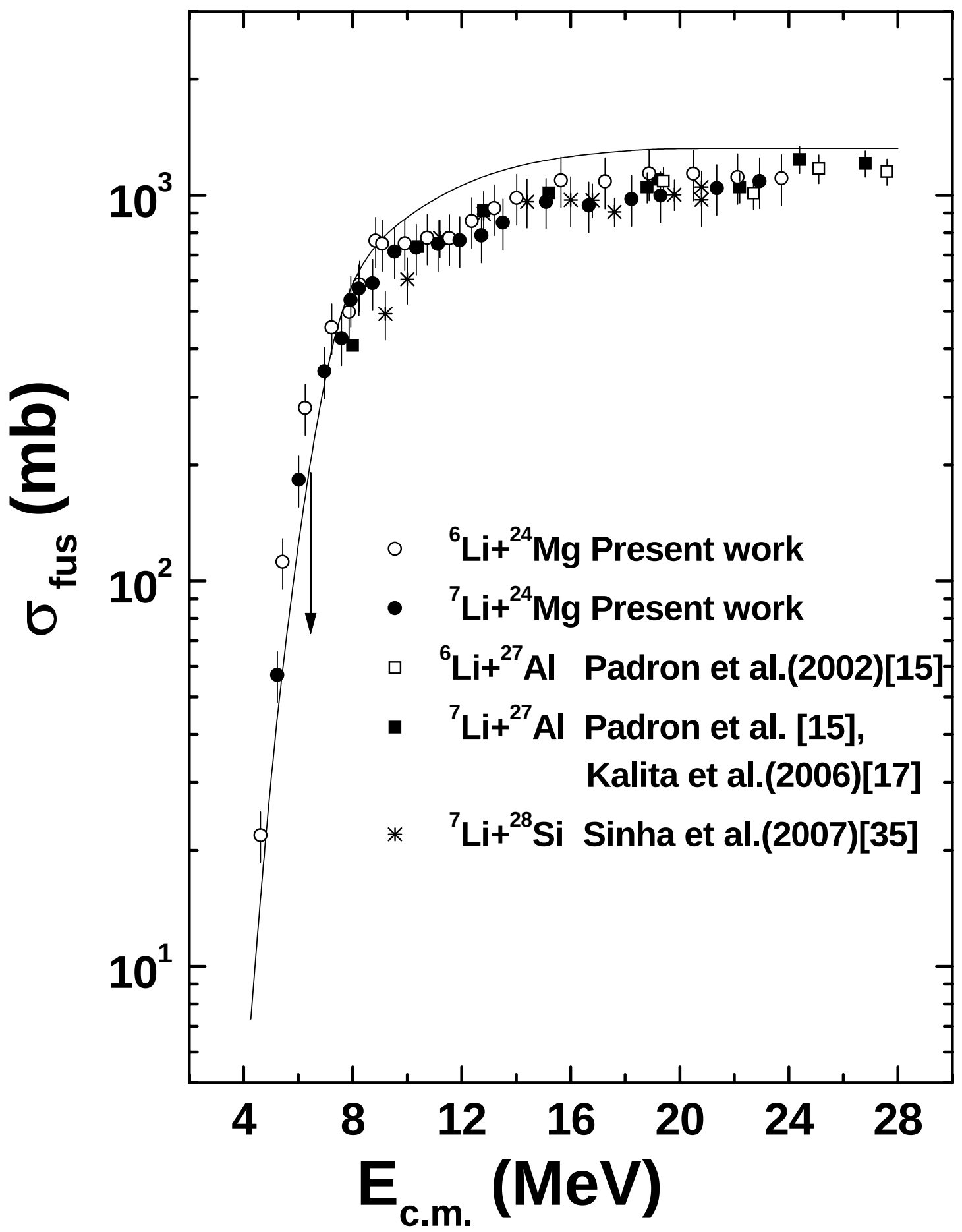

$\sigma_{fus}$ (mb)
$10^3$
$10^2$
$10^1$
4
8
12
16
20
24
28
$E_{c.m.}$ (MeV)
$^6$Li+$^{24}$Mg Present work
$^7$Li+$^{24}$Mg Present work
$^6$Li+$^{27}$Al Padron et al.(2002)[15]
$^7$Li+$^{27}$Al Padron et al. [15], Kalita et al.(2006)[17]
$^7$Li+$^{28}$Si Sinha et al.(2007)[35]

Fig11.

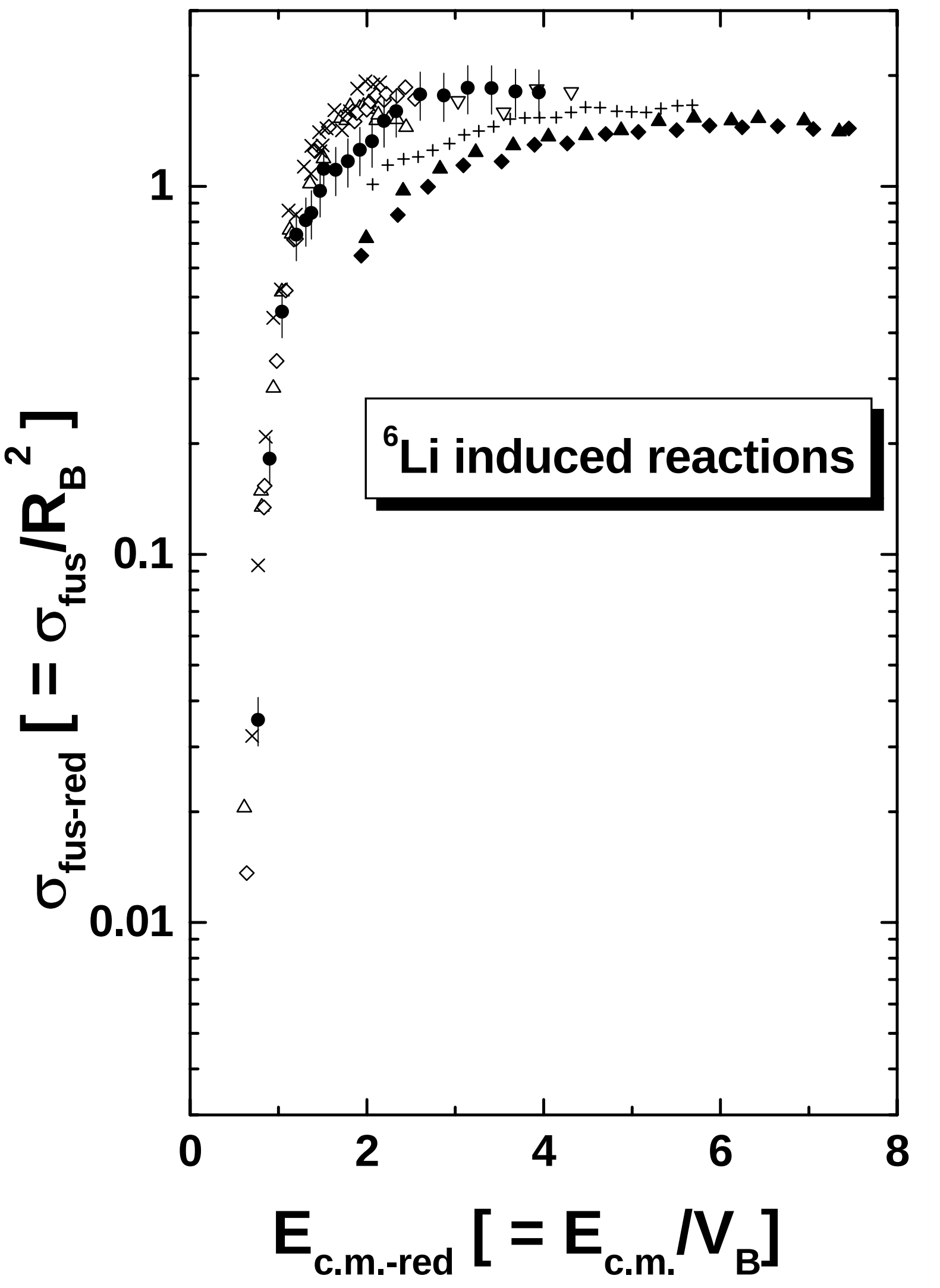

Fig12.

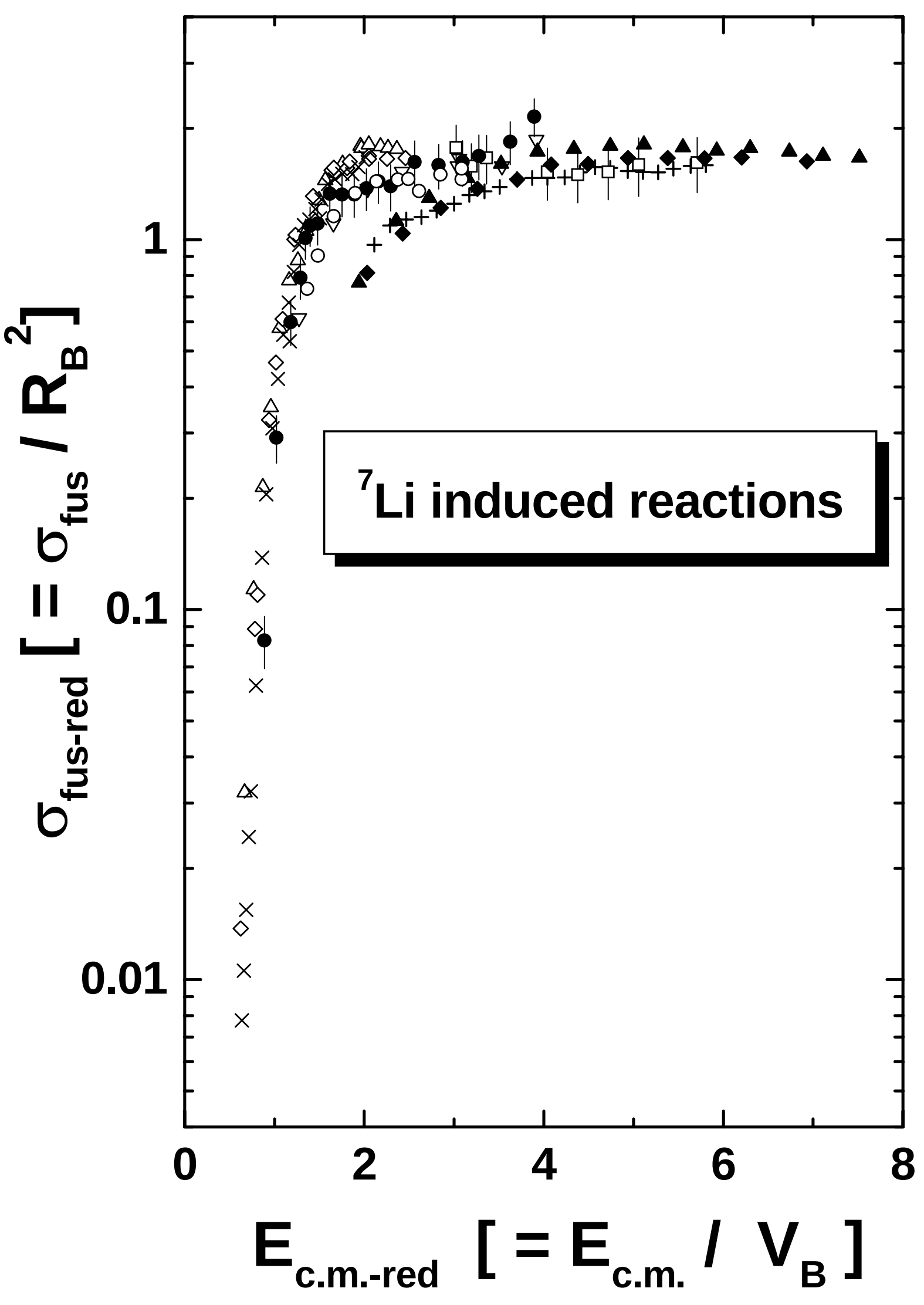

Fig13.